\documentclass[twocolumn,english,pra,superscriptaddress,showpacs,tightenlines]{revtex4}
\usepackage{times}
\usepackage{color}
\usepackage{amsmath}
\usepackage{graphicx}
\usepackage{amssymb}
\usepackage{esint}

\begin{document}

\title{Tunable Electromagnetically Induced Transparency and Absorption with
Dressed Superconducting Qubits}

\author{Hou Ian}

\affiliation{Advanced Science Institute, RIKEN, Wako-shi, Saitama 351-0198, Japan}

\author{Yu-xi Liu}

\affiliation{Advanced Science Institute, RIKEN, Wako-shi, Saitama 351-0198, Japan}

\affiliation{Institute of Microelectronics, Tsinghua University, Beijing 100084,
China}

\affiliation{Tsinghua National Laboratory for Information Science and Technology
(TNList), Tsinghua University, Beijing 100084, China}

\author{Franco Nori}

\affiliation{Advanced Science Institute, RIKEN, Wako-shi, Saitama 351-0198, Japan}

\affiliation{Physics Department, The University of Michigan, Ann Arbor, MI 48109,
USA}
\begin{abstract}
Electromagnetically induced transparency and absorption (EIT and EIA)
are usually demonstrated using three-level atomic systems. In contrast
to the usual case, we theoretically study the EIT and EIA in an equivalent
three-level system: a superconducting two-level system (qubit) dressed
by a single-mode cavity field. In this equivalent system, we find
that both the EIT and the EIA can be tuned by controlling the level-spacing
of the superconducting qubit and hence controlling the dressed system.
This tunability is due to the dressed relaxation and dephasing rates
which vary parametrically with the level-spacing of the original qubit
and thus affect the transition properties of the dressed qubit and
the susceptibility. These dressed relaxation and dephasing rates characterize
the reaction of the dressed qubit to an incident probe field. Using
recent experimental data on superconducting qubits (charge, phase,
and flux qubits) to demonstrate our approach, we show the possibility
of experimentally realizing this proposal. 
\end{abstract}

\pacs{85.25-j, 42.50.Gy, 42.65.An}

\maketitle

\section{Introduction}

\subsection{Electromagnetically induced transparency in optics and superconducting
circuits}

Electromagnetically induced transparency (EIT)~\cite{harris90,fleischhauer05}
manifests spectroscopically the quantized three-level structure of
an atomic medium through its interactions with two semiclassical fields.
It has been widely explored in various contexts (e.g.,~\cite{alzar02,he07,yuan08,bennink01})
since its inception. For example, the EIT effect has been studied
in the context of a two-level atom~\cite{bennink01}, instead of
the usual three-level system. In Ref.~\cite{bennink01}, the energy
levels of the atom are split by a driving optical field into doublets,
equivalent to ac-Stark shifts, and the transparency is realized on
the final four-level system.

\begin{figure}
\includegraphics[bb=56bp 326bp 398bp 832bp,clip,width=8.2cm]{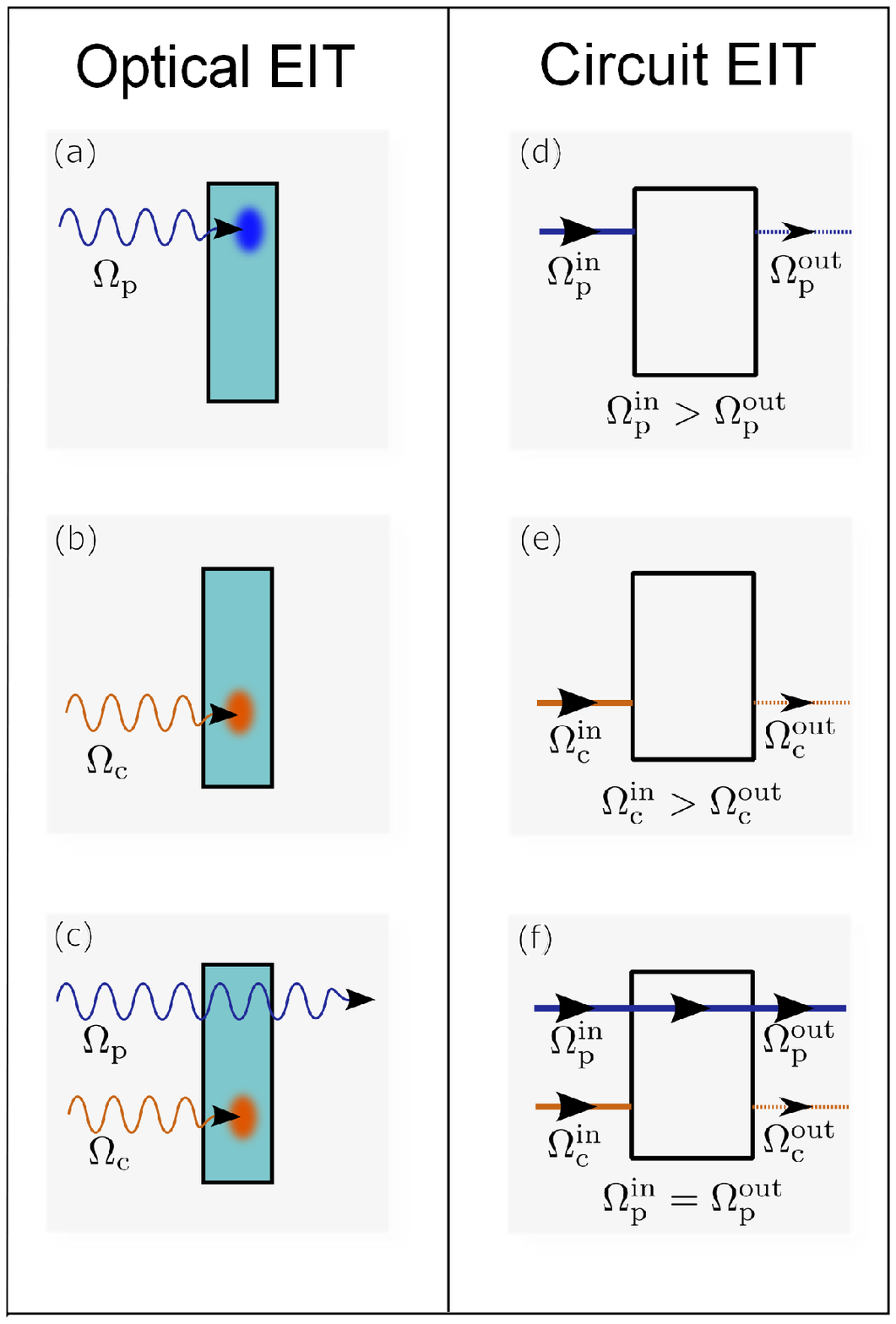}

\includegraphics[bb=55bp 551bp 364bp 755bp,width=6.7cm]{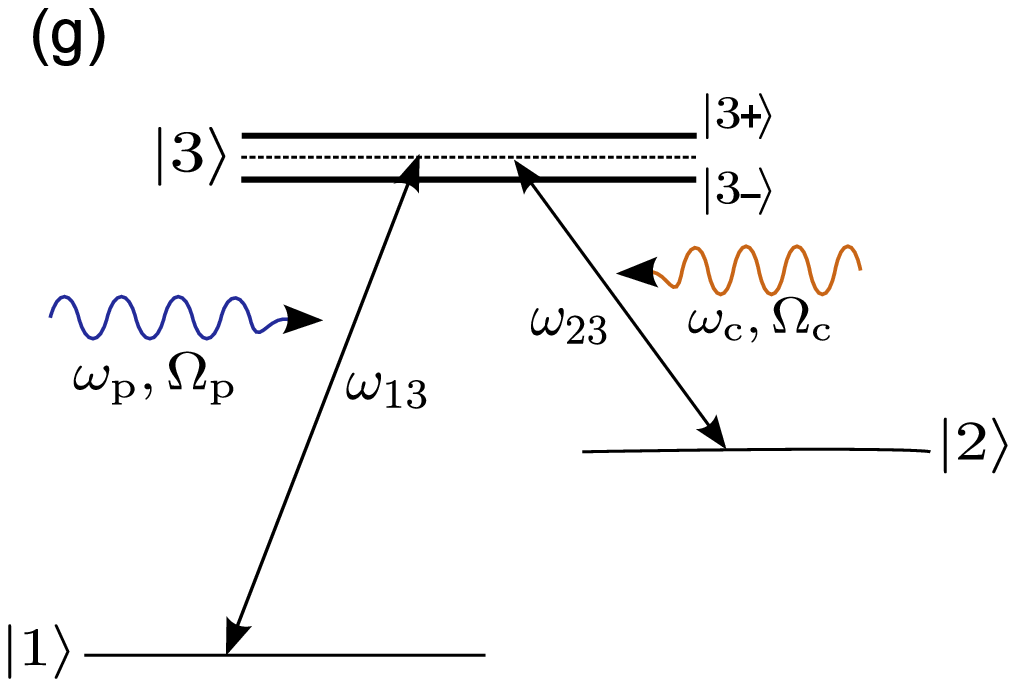}

\caption{(Color online) Schematic diagram illustrating electromagnetically
induced transparency, or EIT. Panel (a) schematically shows a medium
that strongly absorbs a probe light beam of amplitude $\Omega_{\mathrm{p}}$;
(b) shows the same absorption, but for a control light beam of amplitude
$\Omega_{\mathrm{c}}$. However, when both beams are applied simultaneously,
as in (c), then the previously absorbing medium becomes transparent
for the probe beam $\Omega_{\mathrm{p}}$, i.e. transparency is electromagnetically
induced by the control field $\Omega_{\mathrm{c}}$. The superconducting
circuit analog of this phenomenon is schematically shown in (d), (e),
and (f), where a ``box'' schematically represents the circuit. In
(d) a classical voltage or current signal (in the microwave range)
of amplitude $\Omega_{\mathrm{p}}^{\mathrm{in}}$ is fed into the
input of the circuit. The signal is ``absorbed'' in the circuit and
the amplitude $\Omega_{\mathrm{p}}^{\mathrm{out}}$ of the output
signal is less than that of the input. (e) shows the same effect with
another microwave control signal of amplitude $\Omega_{\mathrm{c}}$.
However, when \emph{both} microwave signals are applied simultaneously
to the circuit, as in (f), then the previously ``absorbing medium\textquotedblright{},
now a circuit, becomes \emph{transparent} for the first signal $\Omega_{\mathrm{p}}^{\mathrm{in}}$.
(g) shows a canonical level diagram for a three-level system coupled
to a weak probe field and a strong control field.\label{fig:EIT_illustration}}

\end{figure}

The development of superconducting quantum circuits (SQC) in recent
years have heavily employed concepts from quantum optics, and SQC
have become a testbed of quantum optical phenomena, including EIT
(e.g.,~\cite{sillanpaa09,orlando04,jqyou07}). In one case~\cite{orlando04},
a flux qubit system has even taken advantage of the EIT effect as
a means to probe decoherence by circling the flux qubit with a readout-SQUID
in the SQC. Both the optical version and the SQC version of the EIT
effect are illustrated on the left and right panels, respectively,
of Fig.~\ref{fig:EIT_illustration}, to compare their similarity
and differences. This also serves as a prelude to our discussion of
the tunability of electromagnetically induced transparency and absorption.

For the optical version, on the left of Fig.~\ref{fig:EIT_illustration},
two different kinds of classical electromagnetic waves $\Omega_{\mathrm{p}}(t)=\Omega_{\mathrm{p}}e^{-i\omega_{\mathrm{p}}t}$
and $\Omega_{\mathrm{c}}(t)=\Omega_{\mathrm{c}}\exp^{-i\omega_{\mathrm{c}}t}$,
are shown, for a \textcolor{black}{typically weak probe field $\Omega_{\mathrm{p}}(t)$
and a strong control field $\Omega_{\mathrm{c}}(t)$ (where $\Omega_{\mathrm{c}}\ll\Omega_{\mathrm{p}}$).
These ar}e shown incident on an optical medium in Fig.~\ref{fig:EIT_illustration}(a)
and (b), respectively. The frequencies, $\omega_{\mathrm{p}}$ and
$\omega_{\mathrm{c}}$, of these two fields are resonant with some
energy level spacings of the optical medium, which are typically $\Lambda$-type
three-level atoms, and each of these two fields alone will be absorbed
and cannot travel through the medium. Figure~\ref{fig:EIT_illustration}(g)
shows a canonical level diagram for such three-level atoms, where
the probe field is resonant with the level spacing between $\left|1\right\rangle $
and $\left|3\right\rangle $ and the control field with that between
$\left|2\right\rangle $ and $\left|3\right\rangle $.

When the probe and control fields are simultaneously incident on the
medium, as shown in Fig.~\ref{fig:EIT_illustration}(c), the resonance
between the control field and the medium will render the medium detuned
from the probe field and hence let the probe field travel through
without being absorbed. The level $\left|3\right\rangle $ is driven
out of its original position from $\left|2\right\rangle $ and make
the $\left|1\right\rangle $-to-$\left|3\right\rangle $ spacing detuned
from the probe field frequency in Fig.~\ref{fig:EIT_illustration}(g).
In other words, an absorbing medium becomes transparent to an incident
probe field when a control field is simultaneously applied. More precisely,
the transparency of the probe field can be considered as an effect
of its quantum interference with the control field. The strong coupling
of the medium with the control field perturbs the original level spacings
and provides two excitation pathways of equal probability but opposite
signs to the probe field, indicated by the Autler-Townes levels $\left|3+\right\rangle $
and $\left|3-\right\rangle $ in Fig.~\ref{fig:EIT_illustration}(g).
The resulting signal that exits from the medium is thus a destructive
superposition of two versions of the same signal, hence a destructive
interference and a zero-absorption of the linear susceptibility (see,
e.g., \cite{fleischhauer05} for a comprehensive review). Note that
in the atypical case where both the probe and the control are strongly
coupled to the medium, the interference pattern is severely altered:
in one case, simultaneous transparency for both fields is achieved~\cite{mueller00}
whereas in another, an enhanced absorption of the probe can occur~\cite{wielandy98}.

The SQC version of the optical EIT is illustrated on the right side
of Fig.~\ref{fig:EIT_illustration}, where the optical medium is
replaced by a Josephson-junction multi-level system. In a similar
manner, when the probe and the control signals $\Omega_{\mathrm{p}}^{\mathrm{in}}(t)$
and $\Omega_{\mathrm{c}}^{\mathrm{in}}(t)$, which are either current
or voltage signals in this case, are fed separately into the SQC {[}shown
in Figs.~\ref{fig:EIT_illustration}(d) and (e){]}, the amplitudes
of their outputs $\Omega_{\mathrm{p}}^{\mathrm{out}}(t)$ and $\Omega_{\mathrm{c}}^{\mathrm{out}}(t)$
will be smaller than those of their inputs. This shows that the SQC
separately absorbs the input signals due to resonances between the
input electrical signals and the energy level spacings of the Josephson-junction
device. Nonetheless, a \emph{simultaneous} application of the \emph{two}
input signals would let the probe signal conduct through the circuit
\emph{without} losing energy, while the control signal is mostly absorbed,
as shown in Fig.~\ref{fig:EIT_illustration}(f).

\subsection{Electromagnetically induced absorption}

Two closely-related, but far less studied, optical phenomena are electromagnetically
induced absorption (EIA)~\cite{akulshin98} and switchable dispersion~\cite{goren03},
where the hyperfine structure of the ground state of an atom is used.
The quasi- or near-degenerate levels originate from the same ground
state hyperfine level and have a very small splitting; when the configurations
of their total angular momentum and that of their excited state have
even parity, a $\Delta$-type three-level system with closed cyclic
transitions is formed. Contrary to the odd-parity case, where the
three-level system becomes $\Lambda$-type and can exhibit EIT, the
even-parity configuration makes the multi-level medium absorptive
to the probe field even when it is resonant with the control field.
These relations between optical properties and the parities of SQC
were recently predicted theoretically~\cite{yxliu05} and verified
experimentally~\cite{deppe08} based on selection rules and symmetry-breaking.

\subsection{Tunable transparency and absorption}

The circuit designs in Ref.~\cite{sillanpaa09,orlando04} are based
on the multi-level energy structure of Josephson junction devices.
These designs require advanced measurement techniques where the third
and higher energy levels are often far separated from the bottom two.
We therefore consider in this article an alternative way to construct
a multi-level energy structure in SQC by ``mixing'' a two-level system
with a resonant field, forming a dressed multi-level structure that
is tunable and does not entirely rely on the device characteristics
of the junctions.

Considering the recent progress in studies on superconducting qubits
(e.g.~\cite{you1,you2,you3,you4,you5}) and dressed superconducting
qubits by a single-mode cavity field~\cite{yxliu06,wilson07}, we
study here a dressed three-level qubit equivalent to those in three-level
Josephson junction devices, in order to theoretically realize the
effect of EIT on SQC's. Also motivated by the studies of EIT and EIA
in atomic systems (e.g.,~\cite{harris90,bennink01,akulshin98,goren03}),
we show how the EIT and the EIA phenomena coexist and transmute into
each other on the same dressed SQC.

We will select three energy levels among the multiple levels of the
dressed superconducting qubit. The tunable level spacing of the superconducting
qubit then not only affects the level splitting of the dressed states,
but also affects the relaxation and dephasing rates of the dressed
system. These tunable relaxation and dephasing rates will determine
the system's specific dynamics when coupled to a classical signal
field, effectively making it have either a $\Delta$-type (closed)
or a $\Lambda$-type (open) transition pattern. Therefore, if two
classical electromagnetic fields, the probe field and the control
field, are fed concurrently into the circuit, the complex susceptibility
of the dressed qubit gives an absorption spectrum that \emph{either
dips or peaks} at the zero probe field detuning. The choice of dip
or peak of the spectrum is analytically determined by a biquadratic
equation which is dependent on the qubit level spacing and the environment
temperature through the dressing process. By determining the number
of real roots given by this equation, we can distinguish two effective
regimes of operations: EIT and EIA.

The dip-type and the peak-type spectra corresponds to the EIT and
the EIA effect, respectively, depending on the magnitude of the qubit
level spacing with respect to the eigenfrequency of the resonator
quantum field. Compared to the atomic case, the lower two dressed
states from the superconducting qubit hence act like those hyperfine
levels from the atomic ground state, giving either a closed transition
for the EIA regime and an open transition for the EIT regime. Note
also that these tunable regimes of operations are closely related
to the tunable luminosity suggested by Agarwal \emph{et al.}~\cite{agarwal01},
where the group velocity of light is controlled by a ``knob'' signal
field that couples the metastable states. The tunable metastable-state
coupling given effectively by our dressing process is thus comparable
to that given directly through the knob field.

Our analysis will focus on superconducting quantum circuits that have
a strong coupling between the qubit and the resonator~\cite{zhoulan08,jqliao09},
realized by a coplanar waveguide (CPW) transmission line. These circuits
include combinations of the CPW resonator with either a charge qubit~\cite{you03,wallraff04,wilson07},
a phase qubit~\cite{yxliu04,cleland08,cleland09}, or a flux qubit~\cite{tsai08}.

In Sec.~\ref{sec:undressed_qubit}, we first describe a general theoretical
model and derive the energy spectrum of the dressed multi-level system.
The dressing process is described in Sec.~\ref{sec:dressed_qubit}.
The first-order susceptibility and the dressed relaxation and dephasing
rates among the multiple levels of the qubit-resonator combination
are calculated in Sec.~\ref{sec:susceptibility} using a density
matrix formulation. The determination of the switching between the
transparency and the absorption as well as the discussion of the corresponding
transition patterns are presented in Sec.~\ref{sec:tuning}. In Sec.~\ref{sec:implementations},
we then consider experimentally accessible parameters for different
types of qubits to demonstrate and numerically analyze our theoretical
results. Conclusions are summarized in Sec.~\ref{sec:conclusion}.

\section{Undressed qubit\label{sec:undressed_qubit}}

Our discussion of the superconducting qubit system is independent
of the specific type of qubits (phase qubit, charge qubit, or flux
qubit) employed. Because the two-level structures of these qubits
are commonly described by a general Hamiltonian with a $\sigma_{z}$
term and a $\sigma_{x}$ term~\cite{you1,you2,you3,you4}. To simplify
our discussion, here we consider the qubit in the diagonal basis whose
eigenfrequency $\omega_{q}$ is the root-mean-square of the coefficients
of the $\sigma_{z}$ and $\sigma_{x}$ terms. The CPW resonator that
couples to the qubit, akin to a cavity for photons, is described by
a pair of annihilation and creation operators $a$ and $a^{\dagger}$
with a resonant frequency denoted by $\omega_{0}$. The dipole-field
coupling between the qubit, acting as the dipole, and the energy quantum
within the CPW resonator, acting as the cavity field, is along the
$z$-direction of the qubit in the non-diagonal basis. In the diagonal
basis of the qubit, the Hamiltonian between the qubit and the transmission
line resonator is given by the Jaynes-Cummings model \begin{equation}
H_{\mathrm{cir}}=\omega_{q}\sigma_{z}+\omega_{0}a^{\dagger}a+\eta(a^{\dagger}\sigma_{-}+a\sigma_{+}),\label{eq:cir_ham}\end{equation}
with a rotating-wave approximation and $\hbar=1$, where $\eta$ denotes
the coupling constant between the cavity field and the qubit. Here,
for convenience, we still use $\sigma_{z}$ to denote the qubit operator
in the diagonal basis.

Two signals, the probe signal and the control signal, with traveling
frequencies denoted by $\omega_{\mathrm{p}}$ and $\omega_{\mathrm{c}}$
and Rabi frequencies denoted by $\Omega_{\mathrm{p}}$ and $\Omega_{\mathrm{c}}$,
respectively, are fed into the circuit. These two signals are treated
as classical electromagnetic fields and their interaction Hamiltonian
with the qubit can be written as \begin{equation}
H_{\mathrm{ext}}=\Omega_{\mathrm{p}}e^{i\omega_{\mathrm{p}}t}\sigma_{-}+\Omega_{\mathrm{c}}e^{i\omega_{\mathrm{c}}t}\sigma_{-}+\mathrm{h.c.}\label{eq:ext_ham}\end{equation}
The total system Hamiltonian is then given by \begin{equation}
H=H_{\mathrm{cir}}+H_{\mathrm{ext}}.\label{eq:main_ham}\end{equation}

\begin{figure}
\includegraphics[clip,width=8.5cm]{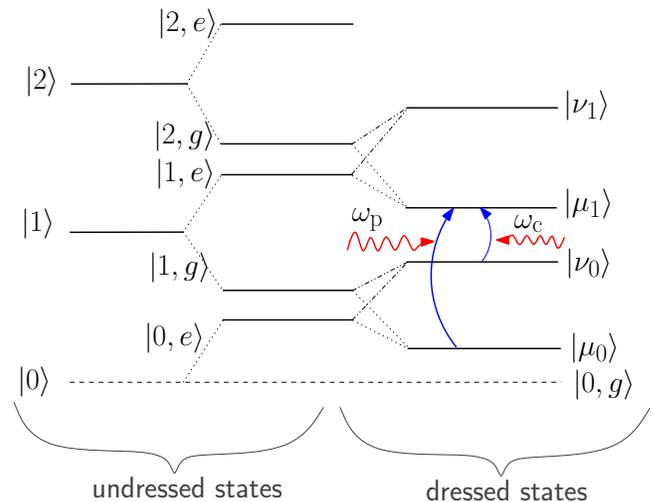}

\caption{Schematic diagram of the ``dressing'' process: the Fock number states
$\left|n\right\rangle $ of a coplanar-waveguide resonator (shown
in the first column) as well as the ground state $\left|g\right\rangle $
and excited state $\left|e\right\rangle $ of a qubit provide the
``undressed'' tensor-product states $\left|n,g\right\rangle $ and
$\left|n,e\right\rangle $ (shown on the middle column). These tensor
states are modified by their mutual interaction, producing the renormalized
or ``dressed'' states $\left|\mu_{n}\right\rangle $ and $\left|\nu_{n}\right\rangle $
(shown on the right). Mathematically, these states correspond to the
eigenvectors of the circuit Hamiltonian in a non-diagonal basis (undressed
states) and the diagonal basis (dressed states), respectively.\label{fig:level_diagram}}

\end{figure}

The energy states of the CPW resonator and the qubit before dressing
are shown as horizontal lines on the left part of the schematic diagram
in Fig.~\ref{fig:level_diagram}. Note that the energy levels $\left|n,g\right\rangle $
and $\left|n,e\right\rangle $ have equal spacings for all $n$: for
fixed probe field and control field, the undressed qubit with a particular
level spacing might not resonate with them. Having been \emph{dressed}
by the CPW resonator, the qubit will exhibit a spectrum with numerous
energy levels spaced in a tunable non-uniform pattern (the dressing
process will be discussed in next section), as shown on the right
part in Fig.~\ref{fig:level_diagram}, providing more possibilities
for matching levels between the dressed qubit and the probe and control
signals.

\section{Dressed Qubit\label{sec:dressed_qubit}}

The dressed qubit Hamiltonian can be derived by rewriting the Jaynes-Cummings
model, which usually describes the atom-photon coupling for the circuit
Hamiltonian $H_{\mathrm{cir}}$ in Eq.~\eqref{eq:cir_ham}. Note
that the set $\{\left|n,e\right\rangle ,\left|n+1,g\right\rangle \}$
spans an invariant subspace $V_{n}$ of $H_{\mathrm{cir}}$, where
$n$ denotes the number of energy quanta in the CPW resonator; while
$e$ and $g$ denote, respectively, the excited and the ground state
of the superconducting qubit. Therefore, the corresponding Hilbert
space, that the Hamiltonian $H_{\mathrm{cir}}$ in Eq.~\eqref{eq:cir_ham}
acts on, can be written as the direct sum

\[
V=\left\{ \left|0,g\right\rangle \right\} \bigoplus_{n=0}V_{n},\]
where the ground state (of the combination of the qubit and the CPW
resonator) does not belong to any invariant subspace. The basis of
each subspace $V_{n}$ can be transformed by rotating an angle\begin{equation}
\theta_{n}=\frac{1}{2}\tan^{-1}\left(\frac{\eta\sqrt{n+1}}{\omega_{q}-\omega_{0}/2}\right)\label{eq:tfm_angle}\end{equation}
such that the circuit Hamiltonian $H_{\mathrm{cir}}$ is diagonalized
in the invariant subspace $V_{n}$ with eigenvalues \begin{equation}
E_{n}=\left(n+\frac{1}{2}\right)\omega_{0}\pm\sqrt{\left(\omega_{q}-\frac{\omega_{0}}{2}\right)^{2}+\eta^{2}(n+1)}\,,\label{eq:eigenvalue}\end{equation}
where the second term constitutes the Rabi splitting for each energy
level $n$. Written in the transformed basis\begin{eqnarray}
\left|\mu_{n}\right\rangle  & = & \cos\theta_{n}\left|n,e\right\rangle -\sin\theta_{n}\left|n+1,g\right\rangle ,\label{eq:mu_n}\\
\left|\nu_{n}\right\rangle  & = & \sin\theta_{n}\left|n,e\right\rangle +\cos\theta_{n}\left|n+1,g\right\rangle \label{eq:nu_n}\end{eqnarray}
and neglecting the ground state $\left|0,g\right\rangle $, the circuit
Hamiltonian~\eqref{eq:cir_ham} can be expressed in its diagonal
``dressed'' form:\begin{equation}
H_{\mathrm{cir}}=\sum_{n}\left[E_{n}^{\mu}\left|\mu_{n}\right\rangle \left\langle \mu_{n}\right|+E_{n}^{\nu}\left|\nu_{n}\right\rangle \left\langle \nu_{n}\right|\right],\label{eq:cir_ham_tfm'd}\end{equation}
where $E_{n}^{\mu}$ ($E_{n}^{\nu}$), associated with the basis vector
$\left|\mu_{n}\right\rangle $ ($\left|\nu_{n}\right\rangle $), corresponds
to the plus (minus) sign of the eigenvalue of Eq.~\eqref{eq:eigenvalue}
at the $n$-th Rabi splitting. The corresponding dressed eigenvectors
are diagrammatically illustrated as the lower (upper) lines on the
right side of Fig.~\ref{fig:level_diagram}. 

Note from Eq.~\eqref{eq:eigenvalue} that the parameter $\omega_{q}$
provides a means to tune the level spacing of the superconducting
qubits, through externally controlling gate voltages, or magnetic
flux, or current~\cite{you1,you2,you3,you4}. Thus the dressed qubit
can exhibit different responses to the probe field signal.

\section{Complex susceptibility\label{sec:susceptibility}}

\subsection{Three-level system}

Originally, the undressed qubit has two levels. When this two-level
system is coupled to a driving CPW resonator, the interaction ``dresses''
the system to have an infinite number of states, instead of two. These
states are shown in Eqs.~\eqref{eq:mu_n}-\eqref{eq:nu_n}. The multi-level
structure of the dressed system gives vast selections of three-level
structures on which the EIT or the EIA effect can be demonstrated.
Before we select three specific levels for our purpose, we shall rewrite
the external part $H_{\mathrm{ext}}$ of the total Hamiltonian, which
we have not discussed so far, in the transformed or dressed basis.

The transformed basis spans the product space of the qubit space and
the resonator space. Considering this, we write the flip-up operator
$\sigma_{+}=I\otimes\left|e\right\rangle \left\langle g\right|=\sum_{n}\left|n,e\right\rangle \left\langle n,g\right|$
as the tensor product of two space bases in $H_{\mathrm{ext}}$. Taking
the inner products of these basis vectors and those of the transformed
basis, we find that, except for the first off-diagonal elements (i.e.,
$\left\langle \mu_{n+1}|\sigma_{+}|\mu_{n}\right\rangle $, $\left\langle \nu_{n+1}|\sigma_{+}|\nu_{n}\right\rangle $,
$\left\langle \mu_{n+1}|\sigma_{+}|\nu_{n}\right\rangle $, and $\left\langle \nu_{n+1}|\sigma_{+}|\mu_{n}\right\rangle $),
all the entries (including the diagonal ones) of the operator $\sigma_{+}$
in the new matrix representation are zero. The $\theta_{n}$-dependent
non-zero matrix elements are all real and, using Eqs.~\eqref{eq:mu_n}-\eqref{eq:nu_n},
the new representation reads \begin{multline}
\sigma_{+}=\sum_{n}\Bigl\{-\cos\theta_{n+1}\sin\theta_{n}\left|\mu_{n+1}\right\rangle \left\langle \mu_{n}\right|+\sin\theta_{n+1}\cos\theta_{n}\\
\times\left|\nu_{n+1}\right\rangle \left\langle \nu_{n}\right|-\sin\theta_{n+1}\sin\theta_{n}\left|\nu_{n+1}\right\rangle \left\langle \mu_{n}\right|\\
+\cos\theta_{n+1}\cos\theta_{n}\left|\mu_{n+1}\right\rangle \left\langle \nu_{n}\right|\Bigr\}.\label{eq:sigma+}\end{multline}
The derivation above applies equally well to the adjoint $\sigma_{-}$.

From Eq.~\eqref{eq:sigma+}, we see that the excitations of the qubit
engage the nearest set of neighboring levels of the CPW resonator
in a way that the transition coefficients depend on the device parameters
of the superconducting qubit. Therefore, from the point of view of
these dressed qubit levels, the process of energy pumping into a resonator
through a mediating qubit~\cite{cleland08} is a laddering of consecutive
level-jumps to the next $(n+1)$ dressed level.

Substituting Eq.~\eqref{eq:sigma+} and its adjoint into Eq.~\eqref{eq:ext_ham},
and also using Eq.~\eqref{eq:cir_ham_tfm'd}, one reaches a total
Hamiltonian expressed completely in the dressed basis. Selecting the
three levels with lowest eigenenergies according to Eq.~\eqref{eq:eigenvalue},
i.e.\[
\left\{ \left|\mu_{0}\right\rangle ,\left|\nu_{0}\right\rangle ,\left|\mu_{1}\right\rangle \right\} ,\]
 a Hamiltonian resembling that of a three-level atom with two associated
dipole-field interactions is obtained \begin{eqnarray}
H_{\Lambda} & = & E_{0}^{\mu}\left|\mu_{0}\right\rangle \left\langle \mu_{0}\right|+E_{0}^{\nu}\left|\nu_{0}\right\rangle \left\langle \nu_{0}\right|+E_{1}^{\mu}\left|\mu_{1}\right\rangle \left\langle \mu_{1}\right|\nonumber \\
 &  & -\Omega_{\mathrm{p}}e^{-i\omega_{\mathrm{p}}t}\cos\theta_{1}\sin\theta_{0}\left|\mu_{1}\right\rangle \left\langle \mu_{0}\right|\nonumber \\
 &  & +\Omega_{\mathrm{c}}e^{-i\omega_{\mathrm{c}}t}\cos\theta_{1}\cos\theta_{0}\left|\mu_{1}\right\rangle \left\langle \nu_{0}\right|+\mathrm{h.c.}\label{eq:ham_Lambda}\end{eqnarray}

The three selected levels, whose transitions are coupled to the classical
probe field of frequency $\omega_{\mathrm{p}}$ and the control field
of frequency $\omega_{\mathrm{c}}$, are shown on the right side of
the level diagram in Fig.~\ref{fig:level_diagram}. Note that when
selecting the levels, the state $\left|0,g\right\rangle $, which
did not participate in the basis transformation, is ignored. The selected
states ($\left|\mu_{0}\right\rangle $, $\left|\nu_{0}\right\rangle $,
and $\left|\mu_{1}\right\rangle $) correspond to the dressed ground
state, metastable state, and excited state, respectively. Each of
the classical fields, $\omega_{\mathrm{p}}$ and $\omega_{\mathrm{c}}$,
can drive transitions between: (i) the ground, metastable, and excited
states, as well as (ii) those between levels of higher energies. We
choose the frequencies of the classical fields such that the probe
field $\omega_{\mathrm{p}}$ (the control field $\omega_{c}$) is
near-resonant with the transition $\left|\mu_{1}\right\rangle \left\langle \mu_{0}\right|$
($\left|\mu_{1}\right\rangle \left\langle \nu_{0}\right|$) for our
discussion of the electromagnetically induced transparency and absorption
effect. The two frequencies $\omega_{\mathrm{p}}$ and $\omega_{\mathrm{c}}$
can be considered to be far-detuned from each other and from other
transitions (including transitions to higher-energy levels). Therefore,
all other types of interactions can be neglected in the Hamiltonian~\eqref{eq:ham_Lambda}.

\subsection{Qubit-resonator interaction}

The coupling coefficients of the three-level system, given in Eq.~\eqref{eq:ham_Lambda},
not only depend on the Rabi frequencies $\Omega_{\mathrm{p}}$ and
$\Omega_{\mathrm{c}}$, but also on the rotation angles $\theta_{0}$
and $\theta_{1}$. That is, the diagonalizing transformation partially
determines the magnitude of the probe and control field couplings.

\begin{figure}
\includegraphics[bb=36bp 203bp 552bp 565bp,clip,width=8.5cm]{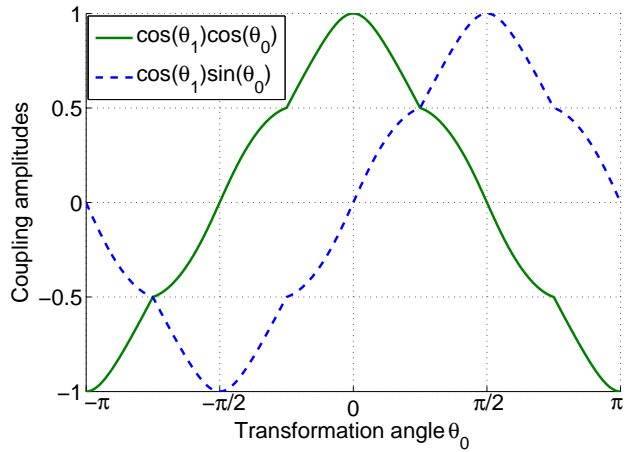}

\caption{Coupling amplitudes of the dressed qubit to the probe field (denoted
by the dashed curve) and to the control field (denoted by the solid
curve).\label{fig:couple_amp}}

\end{figure}

From Eq.~\eqref{eq:tfm_angle}, we observe that the diagonalizing
angles at two different CPW-resonator levels obey the general relation

\[
\frac{\tan2\theta_{n}}{\tan2\theta_{m}}=\frac{\sqrt{n+1}}{\sqrt{m+1}}.\]
 If the levels are such that $m<n$, then $\theta_{n}>\theta_{m}$
in the first quadrant. Applying this relation to $m=0$ and $n=1$,
we find \[
\theta_{1}=\frac{1}{2}\tan^{-1}\left(\sqrt{2}\tan2\theta_{0}\right),\]
which is not an everywhere-differentiable function with respect to
$\theta_{0}$. There exist non-differentiable points at $(l\pi/2+\pi/4)$,
for integer $l$. Figure~\ref{fig:couple_amp} plots the rotation-angle-dependent
factors $(\cos\theta_{1}\sin\theta_{0})$ and $(\cos\theta_{1}\cos\theta_{0})$
in the coupling coefficients of Eq.~\eqref{eq:ham_Lambda}, as functions
of $\theta_{0}$, over one period, within which four non-smooth turning
points can be spotted for each function.

From the first quadrant $(0\leq\theta_{0}\leq\pi/2)$ of Fig.~\ref{fig:couple_amp},
we note that the coupling amplitude between the dressed qubit and
the probe field is monotonically increasing while that of the control
field is monotonically decreasing. The two coupling amplitudes coincide
at $\theta_{0}=\pi/4$. Therefore, the rotation angle can be tuned
in such a way that the coupling strengths between the dressed qubit
and the external signals vary considerably, from $\theta_{0}=0$,
where the coupling to the control field is maximized, to the opposite
limit $\theta_{0}=\pi/2$, where the coupling to the probe field is
maximized.

Coherent trapping, usually discussed in the context of quantum optics,
also occurs here in this superconducting dressed two-level system.
If the initial state of the dressed qubit is prepared with equal populations
in the two lower states with no phase difference, i.e., $\left|\psi(0)\right\rangle =\left(\left|\mu_{0}\right\rangle +\left|\nu_{0}\right\rangle \right)/\sqrt{2}$,
then the interactions with the external signals, as described in Eq.~\eqref{eq:ham_Lambda},
will drive the population at the excited level to be dependent only
on the coupling Rabi frequencies $\Omega_{\mathrm{p}}$ and $\Omega_{\mathrm{c}}$.
In the dressed system, when the qubit level spacing reaches the exact
values\[
\omega_{q}=\frac{\omega_{0}}{2}+\frac{\eta}{\tan\left[2\tan^{-1}(\Omega_{\mathrm{c}}/\Omega_{\mathrm{p}})\right]},\]
the population at the upper lever will remain zero, while those of
the two lower levels stay the same over all time $t$. The phenomenon
of ``trapping'' is then achieved in the sense that the dressed system
will remain in such a configuration with no population in the upper
excited level, even though the classical signals are continually pumped
into the system.

\subsection{Demonstrating EIT via the complex susceptibility}

Using the standard density matrix formalism~\cite{shen}, in this
subsection, we demonstrate the EIT effect through the derivation of
the first-order susceptibility of the dressed qubit. The dressed qubit
acts as a signal-absorbing medium driven by the two external signals
$\omega_{\mathrm{p}}$ and $\omega_{\mathrm{c}}$, as schematically
shown in Fig.~\ref{fig:level_diagram}, similar to the three-level
atom with the optical fields given in Fig.~\ref{fig:EIT_illustration}(g).
We will use the matrix element notation $\rho_{\alpha\beta}=\left|\alpha\right\rangle \left\langle \beta\right|$,
where $\alpha$ and $\beta$ can be one of the symbols $\mu,\nu$
for the lower levels $\left|\mu_{0}\right\rangle ,\left|\nu_{0}\right\rangle $
and 1 for the excited level $\left|\mu_{1}\right\rangle $; $\rho_{\alpha\beta}$
denotes level populations when $\alpha=\beta$, or transition amplitudes
otherwise. We shall also use the shorthands for the coupling coefficients:\begin{eqnarray*}
\zeta_{\mathrm{p}}(t) & = & -\Omega_{\mathrm{p}}e^{-i\omega_{\mathrm{p}}t}\cos\theta_{1}\sin\theta_{0},\\
\zeta_{\mathrm{c}}(t) & = & \Omega_{\mathrm{c}}e^{-i\omega_{\mathrm{c}}t}\cos\theta_{1}\cos\theta_{0}.\end{eqnarray*}

The matrix elements thence evolves according to Schroedinger's equation
with respect to the Hamiltonian in Eq.~\eqref{eq:ham_Lambda}, analogous
to the optical Bloch equations,\begin{subequations} \begin{align}
\dot{\rho}_{\mu\mu} & =-\Gamma_{\mu}\rho_{\mu\mu}-i\zeta_{\text{p}}\rho_{1\mu}+i\zeta_{\mathrm{p}}^{\ast}\rho_{\mu1},\\
\dot{\rho}_{\nu\nu} & =-\Gamma_{\nu}\rho_{\nu\nu}-i\zeta_{\mathrm{c}}\rho_{1\nu}+i\zeta_{\mathrm{c}}^{\ast}\rho_{\nu1},\\
\dot{\rho}_{11} & =-\Gamma_{1}\rho_{11}+i\zeta_{\mathrm{p}}\rho_{1\mu}+i\zeta_{\mathrm{c}}\rho_{1\nu}-i\zeta_{\mathrm{p}}^{\ast}\rho_{\mu1}-i\zeta_{\mathrm{c}}^{\ast}\rho_{\nu1},\\
\dot{\rho}_{\mu1} & =-\left[i(E_{0}^{\mu}-E_{1}^{\mu})+\gamma_{\mu1}\right]\rho_{\mu1}-i\zeta_{\mathrm{p}}(\rho_{11}-\rho_{\mu\mu})+i\zeta_{\mathrm{c}}\rho_{\mu\nu},\label{eq:EoM_mu1}\\
\dot{\rho}_{\nu1} & =-\left[i(E_{0}^{\nu}-E_{1}^{\mu})+\gamma_{\nu1}\right]\rho_{\nu1}-i\zeta_{\mathrm{c}}(\rho_{11}-\rho_{\nu\nu})+i\zeta_{\mathrm{p}}\rho_{\mu\nu}^{\ast},\\
\dot{\rho}_{\mu\nu} & =-\left[i(E_{0}^{\mu}-E_{0}^{\nu})+\gamma_{\mu\nu}\right]\rho_{\mu\nu}-i\zeta_{\mathrm{p}}\rho_{\nu1}^{\ast}+i\zeta_{\mathrm{c}}^{\ast}\rho_{\mu1}.\label{eq:EoM_munu}\end{align}
\end{subequations}In the equations above, we have added phenomenologically
the relaxation rate $\Gamma_{\mu}$ ($\Gamma_{\nu}$, $\Gamma_{1}$)
for the level $\left|\mu_{0}\right\rangle $ ($\left|\nu_{0}\right\rangle $,
$\left|\mu_{1}\right\rangle $) as well as the dephasing rate $\gamma_{\mu1}$
($\gamma_{\nu1}$, $\gamma_{\mu\nu}$) for the transition $\left|\mu_{0}\right\rangle \left\langle \mu_{1}\right|$
($\left|\nu_{0}\right\rangle \left\langle \mu_{1}\right|$, $\left|\mu_{0}\right\rangle \left\langle \nu_{0}\right|$).
Note that the subscripts $\mu$, $\nu$, and 1 here refer to the levels
$\left|\mu_{0}\right\rangle $, $\left|\nu_{0}\right\rangle $ and
$\left|\mu_{1}\right\rangle $, respectively, to simplify the notation.
This system of equations is homogeneous, which gives a zero steady-state
solution, i.e. the system's thermal equilibrium state, when the coefficient
matrix is nondegenerate. Therefore, we can assign, without loss of
generality, the population under consideration entirely to the ground
state, i.e. $\rho_{\mu\mu}^{(0)}=1$ and $\rho_{\nu\nu}^{(0)}=\rho_{11}^{(0)}=0$,
and the polarization $\left\langle \mathbf{P}^{(0)}\right\rangle $
(dipole moment) amongst the three energy levels at the zeroth-order
expansion to zero value, i.e. $\rho_{\mu1}^{(0)}=\rho_{\nu1}^{(0)}=\rho_{\mu\nu}^{(0)}=0$.

Since we are concerned with the dispersion and absorption spectrum
of the dressed qubit, only the first-order perturbative expansion
of the density matrix elements in Eqs.~\eqref{eq:EoM_mu1}-\eqref{eq:EoM_munu}
are needed. Substituting the steady-state solution above into these
two equations and removing the time dependences of the coefficients
in the rotating frame of reference $\rho_{\mu1}^{(1)}\to\rho_{\mu1}^{(1)}\exp\{-i\omega_{\mathrm{p}}t\}$,
$\rho_{\mu\nu}^{(1)}\to\rho_{\mu\nu}^{(1)}\exp\left\{ -i\left(\omega_{\mathrm{p}}+E_{0}^{\nu}-E_{1}^{\mu}\right)\right\} $,
we have\begin{align}
\dot{\rho}_{\mu1}^{(1)}= & -(i\Delta+\gamma_{\mu1})\rho_{\mu1}^{(1)}+i\Omega_{\mathrm{c}}\cos\theta_{1}\cos\theta_{0}\rho_{\mu\nu}^{(1)}\nonumber \\
 & -i\Omega_{\mathrm{p}}\cos\theta_{1}\sin\theta_{0},\label{eq:EoM_mu1_1st}\\
\dot{\rho}_{\mu\nu}^{(1)}= & -(i\Delta+\gamma_{\mu\nu})\rho_{\mu\nu}^{(1)}+i\Omega_{\mathrm{c}}\cos\theta_{1}\cos\theta_{0}\rho_{\mu1}^{(1)},\label{eq:EoM_munu_1st}\end{align}
where \[
\Delta=(E_{1}^{\mu}-E_{0}^{\mu})-\omega_{\mathrm{p}}\]
is the detuning between the probe field and the level spacing of the
ground state and the excited state. The control field frequency $\omega_{\mathrm{c}}$
is assumed to match the level spacing between the metastable state
$\left|\nu_{0}\right\rangle $ and the excited state $\left|\mu_{1}\right\rangle $:
$\omega_{\mathrm{c}}=E_{1}^{\mu}-E_{0}^{\nu}$.

The strength of the probe field is related to the polarization of
the medium through the relation\[
\epsilon_{0}\chi^{(1)}\Omega_{\mathrm{p}}=|\mathbf{d}_{\mu1}|^{2}\rho_{\mu1}\]
where $\epsilon_{0}$ denotes the vacuum permittivity and $\mathbf{d}_{\mu1}$
the qubit dipole moment. Combining this relation with Eq.~\eqref{eq:EoM_mu1_1st}
and Eq.~\eqref{eq:EoM_munu_1st} in steady state, we can find the
first-order susceptibility \[
\chi^{(1)}=\chi'+i\chi''\]
decomposed in a real part\begin{multline}
\chi'=Z\Delta\Bigl[\Delta^{2}-\gamma_{\mu1}\gamma_{\mu\nu}-(\Omega_{\mathrm{c}}\cos\theta_{1}\cos\theta_{0})^{2}\\
+\gamma_{\mu\nu}(\gamma_{\mu1}+\gamma_{\mu\nu})\Bigl]\label{eq:scpt_re}\end{multline}
 and an imaginary part\begin{equation}
\chi''=Z\left[\gamma_{\mu1}\Delta^{2}+\gamma_{\mu1}\gamma_{\mu\nu}^{2}+\gamma_{\mu\nu}(\Omega_{\mathrm{c}}\cos\theta_{1}\cos\theta_{0})^{2}\right]\label{eq:scpt_im}\end{equation}
 where the common factor is\begin{multline*}
Z=\frac{|\mathbf{d}_{\mu1}|^{2}}{\epsilon_{0}}\cos\theta_{1}\sin\theta_{0}\biggl\{\Delta^{2}(\gamma_{\mu1}+\gamma_{\mu\nu})^{2}\\
+\left[\Delta^{2}-\gamma_{\mu1}\gamma_{\mu\nu}-(\Omega_{\mathrm{c}}\cos\theta_{1}\cos\theta_{0})^{2}\right]^{2}\biggr\}^{-1}.\end{multline*}
Note that the dependences of Eqs.~\eqref{eq:scpt_re}-\eqref{eq:scpt_im}
on the detuning $\Delta$ are similar to those of the susceptibilities
of regular three-level atomic systems. The difference here is the
extra dependence on the qubit level-spacing $\omega_{q}$ through
the transformation angles $\theta_{1}$ and $\theta_{0}$. First,
the Rabi frequency $(\Omega_{\mathrm{c}}\cos\theta_{1}\cos\theta_{0})$,
which describes the coupling strength of the control signal to the
dressed qubit, has an explicit dependence on the angles $\theta_{1}$
and $\theta_{0}$. Second, the relaxation rates $\gamma_{\mu1}$ and
$\gamma_{\mu\nu}$ have implicit dependences on these angles. These
cosinusoidal dependences will vary the magnitude of every term and
will hence vary the functional behaviors of the two parts $\chi'$
and $\chi''$ of the susceptibility. In the following sections, we
will see that these dependences are crucial to the tunability of the
dressed qubit to become either transparent or absorptive.

\section{Decay rates between the dressed levels}

\subsection{Overview of recent theories}

In this section, we study how the relaxation rates $\gamma_{\mu1}$
and $\gamma_{\mu\nu}$ in Eqs.~\eqref{eq:scpt_re}-\eqref{eq:scpt_im}
and other dephasing rates are implicitly dependent on the qubit level
spacing $\omega_{q}$ and other system parameters. In particular,
we will determine how these decay rates are affected by the dressing
process, i.e., after the qubit levels are rotated into the new dressed
basis.

The decay of an undressed qubit induced by an environmental coupling
has been extensively studied in Refs.~\cite{ithier05} and \cite{hauss08}
through a first-order perturbation model based on the Bloch and Redfield's
analysis~\cite{bloch57} on what was called a relaxation or distribution
matrix. The relaxation process induced by the environment has many
implications on multi-level systems, including the lasing action of
superconducting circuits~\cite{ashhab09}.

The recent article by Wilson \emph{et al.}~\cite{wilson09} is the
first study that took a step further to deal with the dressing effect
on relaxation and dephasing, in which the resonator field couples
to the $z$-direction of the qubit and diagonalizing in a displaced
basis results in a splitting of levels that has a Bessel-function
dependence on the quantum number $n$. The approach here with the
rotating wave approximation, in contrast, has a linear dependence
of the splitting on $n$ (Cf. Eq.~\eqref{eq:eigenvalue}). This number
$n$ determines both the thermal distribution of energy quanta in
the CPW resonator and the rotation angles of the levels in each invariant
subspace, for which we expect to see a two-fold dependence of the
relaxation rates on the level $n$. Moreover, the environmental temperature
$T$ determines the CPW-resonator's thermal distribution, as well
as the magnitude of the fluctuations induced by the thermal coupling
of the non-dressed qubit. Therefore, the dressed qubit has a two-fold
dependence on the temperature $T$.

We will study each aspect of this dependence separately in the following
subsections. The direct fluctuations introduced by the thermal coupling
before the dressing process will be discussed first. The indirect
dependence through the resonator's thermal distribution after the
dressing process will follow, where the implicit dependence on the
qubit level spacing $\omega_{q}$ is also introduced.

\subsection{Relaxation and dephasing before dressing}

The two steps of computing the dressed decay rates can be separated
by first assuming that the system has reached a thermal equilibrium
and the two-level qubit provides the only source of noise (i.e., the
resonator has no thermal fluctuations). Following the standard methodology
(e.g. in Ref.~\cite{ithier05}), we describe the thermal environment
by a single quantum variable $X$, which interacts with the qubit
along all three directions through the Hamiltonian $H_{I}=(c_{x}\sigma_{x}+c_{y}\sigma_{y}+c_{z}\sigma_{z})X$.
Among the interaction coefficients, those of the $\sigma_{x}$- and
$\sigma_{y}$-directions affect the population decay between two levels,
i.e. they contribute a total relaxation rate \[
r_{1}=r_{\downarrow}+r_{\uparrow}\]
which consists of a ``down'' relaxation rate $r_{\downarrow}$ and
an ``up'' relaxation rate $r_{\uparrow}$: \begin{align*}
r_{\downarrow} & =|c_{x}+ic_{y}|^{2}S_{X}(\omega)/4,\\
r_{\uparrow} & =|c_{x}+ic_{y}|^{2}S_{X}(-\omega)/4.\end{align*}
The $S_{X}$ shown in the expressions above indicate the power spectrum
of the heat bath and its value has an exponential dependence on the
environment temperature $T$, \[
S_{X}(\omega)=\frac{1}{2\pi}\int_{-\infty}^{\infty}\mathrm{d}\omega'\left\langle X(\omega)X(\omega')\right\rangle =\frac{R\omega}{2\pi}\coth\left(\frac{\omega}{2k_{B}T}\right),\]
where $R$ denotes a nominal resistance.

The $\sigma_{z}$-direction of the interaction coefficients contributes
a part, $r_{\varphi}$, of the total dephasing rate, which is known
as the pure dephasing part \[
r_{\varphi}=|c_{z}|^{2}S_{X}(0)/2.\]
The interactions along $\sigma_{x}$- and $\sigma_{y}$-directions
contribute the other part, $r_{1}$, of the dephasing, which make
the total dephasing rate of the qubit sum up to\[
r_{2}=\frac{r_{1}}{2}+r_{\varphi}.\]

\subsection{Relaxation and dephasing after dressing}

Considering the level mixings due to the unitary transformations in
Eqs.~\eqref{eq:mu_n} and \eqref{eq:nu_n}, the fluctuations of the
dressed levels become the superpositions of the fluctuations originated
from the states of the undressed qubit.

The density matrix element for the lowest level of the dressed qubit,
expanded in the undressed basis, reads\begin{multline*}
\left|\mu_{0}\right\rangle \left\langle \mu_{0}\right|=\cos^{2}\theta_{0}\left|0,e\right\rangle \left\langle 0,e\right|+\sin^{2}\theta_{0}\left|1,g\right\rangle \left\langle 1,g\right|\\
-\cos\theta_{0}\sin\theta_{0}\left[\left|0,e\right\rangle \left\langle 1,g\right|+\mathrm{h.c.}\right].\end{multline*}
We see that this matrix element has the diagonal parts and the off-diagonal
parts in the undressed basis. To simplify the formulation, we assume
that the dressed qubit has reached a thermal equilibrium and ignore
the system relaxation due to the energy exchange between the CPW resonator
and the qubit, that is, the off-diagonal parts. Then by tracing out
the subspace of the resonator part by assuming its energy quanta has
a Boltzmann distribution, we can obtain the reduced density matrix
element\[
\rho'_{\mu\mu}=(1-e^{-\beta\omega_{0}})\left[\cos^{2}\theta_{0}\left|e\right\rangle \left\langle e\right|+e^{-\beta\omega_{0}}\sin^{2}\theta_{0}\left|g\right\rangle \left\langle g\right|\right],\]
where $\beta=1/k_{B}T$ denotes the inverse temperature. Considering
that the two levels of the undressed qubit contributes equally to
the up and down relaxations, we arrive at the relaxation rate $\Gamma_{\mu}$
of the lowest energy level $\left|\mu_{0}\right\rangle $ of the dressed
qubit\begin{subequations}

\begin{equation}
\Gamma_{\mu}=\frac{|c_{x}+ic_{y}|^{2}S_{X}(\omega)}{4}(1-e^{-\beta\omega_{0}})\left[\cos^{2}\theta_{0}+e^{-\beta\omega_{0}}\sin^{2}\theta_{0}\right].\label{eq:Gamma_mu}\end{equation}
Using similar steps, we can derive the relaxation rates ($\Gamma_{\nu}$
and $\Gamma_{1}$) of the other energy levels ($\left|\nu_{0}\right\rangle $
and $\left|\mu_{1}\right\rangle $) as well as their dephasing rates
($\gamma_{\mu1}$, $\gamma_{\nu1}$, and $\gamma_{\mu\nu}$):\begin{align}
\Gamma_{\nu} & =\frac{|c_{x}+ic_{y}|^{2}S_{X}(\omega)}{4}(1-e^{-\beta\omega_{0}})\left[\sin^{2}\theta_{0}+e^{-\beta\omega_{0}}\cos^{2}\theta_{0}\right],\\
\Gamma_{1} & =\frac{|c_{x}+ic_{y}|^{2}S_{X}(\omega)}{4}e^{-\beta\omega_{0}}(1-e^{-\beta\omega_{0}})\nonumber \\
 & \qquad\qquad\qquad\times\left[\cos^{2}\theta_{1}+e^{-\beta\omega_{0}}\sin^{2}\theta_{1}\right],\\
\gamma_{\mu1} & =-\frac{|c_{z}|^{2}S_{X}(0)}{2}e^{-\beta\omega_{0}}(1-e^{-\beta\omega_{0}})\sin\theta_{0}\cos\theta_{1},\label{eq:gamma_mu1}\\
\gamma_{\nu1} & =\frac{|c_{z}|^{2}S_{X}(0)}{2}e^{-\beta\omega_{0}}(1-e^{-\beta\omega_{0}})\cos\theta_{0}\cos\theta_{1},\\
\gamma_{\mu\nu} & =\frac{|c_{x}+ic_{y}|^{2}S_{X}(\omega)}{4}(1-e^{-\beta\omega_{0}})^{2}\cos\theta_{0}\sin\theta_{0}.\label{eq:gamma_munu}\end{align}
\end{subequations}

We note from Eq.~\eqref{eq:gamma_munu} that the dephasing rate $\gamma_{\mu\nu}$
between the split states $\left|\mu_{0}\right\rangle $ and $\left|\nu_{0}\right\rangle $,
which share the same number of energy quanta, originates from the
relaxation rates of the undressed qubit levels $\left|e\right\rangle $
and $\left|g\right\rangle $. Therefore, in transforming the basis
for diagonalization, the roles of relaxation and dephasing have exchanged.

The total relaxation rate among the three dressed levels becomes now
\[
\Gamma=\Gamma_{\mu}+\Gamma_{\nu}+\Gamma_{1}\]
which has a two-fold dependence on the environment temperature $T$:
through the noise spectrum $S_{X}(\omega)$ and through the resonator
distribution $\exp(-\beta\omega_{0})$. The total relaxation rate
$\Gamma$ is also tunable by the qubit level spacing $\omega_{q}$
through the transformation angle $\theta_{0}$.

\section{Tuning the electromagnetically induced transparency and absorption\label{sec:tuning}}

\subsection{Local extrema of the susceptibility}

The qubit, together with the CPW resonator, acts as a nonlinear medium
of electron propagation in the quantum circuit. This nonlinear medium
gives the rotation-angle dependent and relaxation-rate dependent responses
of both the dispersion and the absorption to the incident probe signal
$\Omega_{\mathrm{p}}e^{-i\omega_{\mathrm{p}}t}$. These two responses
are quantified by the real and the imaginary parts of the susceptibility,
respectively, in Eqs.~\eqref{eq:scpt_re}-\eqref{eq:scpt_im}. Since
the relaxation rates $\gamma_{\mu1},\gamma_{\mu\nu}$ themselves depend
on the rotation angles $\theta_{0},\theta_{1}$ (Cf. Eqs.~\eqref{eq:gamma_mu1}-\eqref{eq:gamma_munu}),
which in turn depend on the qubit level spacing $\omega_{q}$ (Cf.
Eq.~\eqref{eq:tfm_angle}), the spacing $\omega_{q}$ tunes the dispersion
and absorption spectra by controlling the Josephson coupling energy.

The absorption spectrum as a function of the detuning $\Delta$ between
the probe signal and the dressed level spacing $(E_{0}^{\mu}-E_{1}^{\mu})$
is particularly interesting with its number of maxima depending on
the spacing $\omega_{q}$. The derivative of $\chi''$ with respect
to $\Delta$ is the product of $\Delta$ and a quartic expression
of $\Delta$. Therefore, the absorption spectrum always takes an extremal
value at the zero root $\Delta_{0}=0$. The quartic expression is
actually biquadratic, whose two roots are given by\begin{multline}
\Delta_{\pm}=\pm\frac{1}{\sqrt{\gamma_{\mu1}}}\Biggl\{-\gamma_{\mu\nu}\left[\zeta_{\mathrm{c}}^{2}(0)+\gamma_{\mu1}\gamma_{\mu\nu}\right]\\
+(\gamma_{\mu1}+\gamma_{\mu\nu})\zeta_{\mathrm{c}}(0)\sqrt{\zeta_{\mathrm{c}}^{2}(0)+\gamma_{\mu1}\gamma_{\mu\nu}}\Biggr\}^{1/2},\label{eq:Delta_root}\end{multline}
where the other two roots that are associated with the negative sign
of the second term are omitted since the detuning $\Delta$ can only
admit real values. The two admissible roots coincide and meet the
zero root, i.e. $\Delta_{\pm}=\Delta_{0}=0$, when\begin{equation}
(\Omega_{\mathrm{c}}\cos\theta_{1}\cos\theta_{0})^{2}+\gamma_{\mu1}\gamma_{\mu\nu}=0.\label{eq:crit_cond}\end{equation}
 That is, at near resonance $\omega_{q}\approx\omega_{0}/2$, when
the qubit level spacing reaches the critical values\begin{multline}
\lambda_{\mathrm{C},\pm}=\frac{1}{2}\Biggl\{\omega_{0}\pm\frac{\eta}{F^{2}-\sqrt{2}}\left[(\sqrt{2}+1)F^{2}+2\sqrt{2}\right.\\
\left.-F\sqrt{(\sqrt{2}-1)^{2}F^{2}+16+8\sqrt{2}}\right]\Biggr\},\label{eq:crit_Omega}\end{multline}
 where\[
F(\Omega_{\mathrm{c}},T)=\frac{8\Omega_{\mathrm{c}}^{2}e^{\beta\omega_{0}}(1-e^{-\beta\omega_{0}})^{-3}}{|c_{x}+ic_{y}|^{2}|c_{z}|^{2}S_{X}(0)S_{X}(\omega)}\]
is a control field amplitude-dependent and temperature-dependent factor.
We can also check that the condition, $(\Omega_{\mathrm{c}}\cos\theta_{1}\cos\theta_{0})^{2}+\gamma_{\mu1}\gamma_{\mu\nu}>0$,
has already guaranteed the right hand side of Eq.~\eqref{eq:Delta_root}
to be greater than zero (See Appendix~A).

Consequently, inside the critical range $\lambda_{\mathrm{C},-}<\omega_{q}<\lambda_{\mathrm{C},+}$,
the only admissible root of $\mathrm{d}\chi''/\mathrm{d}\Delta$ occurs
at the zero point, and from the second-order derivative, it can be
seen that this root corresponds to a local maximum. In the opposite
case, when the qubit level spacing $\omega_{q}$ is tuned outside
the critical range, two new extrema arise in the absorption spectrum,
giving a total of three turning points in the absorption curve. Because
$\chi''$ is an even function of $\Delta$, the original peak point
at zero detuning splits symmetrically about the origin, creating two
symmetric peaks whose distance $(\Delta_{+}-\Delta_{-})$ is extended
when $\omega_{q}$ is tuned away from the critical values $\lambda_{\mathrm{C},\pm}$
at either end; whereas the original peak \[
\chi''\biggl|_{\Delta_{0}}\propto\left[1+\Omega_{\mathrm{c}}^{2}\frac{\cos\theta_{1}\cos\theta_{0}}{C\sin^{2}\theta_{0}}\right]^{-1}\]
 itself, where $C$ denotes some constant, starts to dip from a maximum
value to a local minimum. This local minimum tends to zero when the
spacing $\omega_{q}$ is tuned away from its resonant value $\omega_{0}/2$
and the amplitude $\Omega_{\mathrm{c}}$ of the control field is increased.

This ``peaking-to-dipping'' transition (or the increase of the number
of extrema) indicates the \emph{switching of the dressed qubit from
being transparent to being absorptive} to the probe field. The magnitude
of the qubit level spacing $\omega_{q}$ with respect to the frequency
$\omega_{0}$ of the resonator field determines the nature of the
dressed medium. The critical condition of Eq.~\eqref{eq:crit_cond}
indicates the competition between the population pumping to the excited
level and the spontaneous relaxations to the ground state. The transparency
effect is present only when the coherent pumping is sufficiently strong
to overcome the relaxation; otherwise, the probe signal is trapped
and the dressed medium becomes absorbing.

While the imaginary part of the susceptibility is an even function
of $\Delta$, the real part is odd and one-order higher in $\Delta$
than the imaginary part. The dispersion spectrum similarly admits
multiple local extrema, though the dispersion is always zero at $\Delta=0$
for any value of qubit level spacing $\omega_{q}$ because of the
odd symmetry about the origin. However, when sweeping $\omega_{q}$
across the resonance point $\omega_{0}/2$, the spectrum is inverted,
i.e. the dispersion is switched from positive to negative or vice
versa.

\subsection{Near-degeneracy and the switching from closed to open transitions}

The phenomena discussed in the last subsection are somewhat analogous
to those presented in Refs.~\cite{akulshin98,goren03}, so here we
give a physical interpretation of the tuning from transparency to
absorption using the terminology of atomic physics.

The three dressed states $\left|\mu_{0}\right\rangle $, $\left|\nu_{0}\right\rangle $
and $\left|\mu_{1}\right\rangle $ that we have selected as the basis
of the three-level system have tunable level spacings based on the
transformation angles $\theta_{0}$ and $\theta_{1}$ which are defined
by the detuning between the parameters $\omega_{q}$ and $\omega_{0}$.
After coupling to the strong control field, the two dressed states
$\left|\nu_{0}\right\rangle $ and $\left|\mu_{1}\right\rangle $
will have line-broadening. The metastable state $\left|\nu_{0}\right\rangle $
in particular might be so broadened that it overlaps with the ground
state $\left|\mu_{0}\right\rangle $. Such a case most easily occurs
at resonance with $\omega_{q}=\omega_{0}/2$, where the spacing between
the lower two states $\left|\mu_{0}\right\rangle $ and $\left|\nu_{0}\right\rangle $
is minimized to $2\eta$ (Cf. Eq.~\eqref{eq:eigenvalue}).

When the overlap occurs, the three-level system becomes effectively
two-level. The lower two levels $\left|\mu_{0}\right\rangle $ and
$\left|\nu_{0}\right\rangle $ become quasi- or near-degenerate; more
precisely, they degenerate into a single ground state and differ from
each other only as hyperfine levels of the common ground state. The
minimal amplitude $\Omega_{\mathrm{c}}$ of the coupling needed for
overlapping can be roughly estimated using a first-order perturbative
expansion. Considering the spacing between the metastable state $\left|\nu_{0}\right\rangle $
and the excited state $\left|\mu_{1}\right\rangle $ to be $E_{1}^{\mu}-E_{0}^{\nu}=\omega_{0}-(\sqrt{2}+1)\eta$,
we find the level shift of $\left|\nu_{0}\right\rangle $ up to first
order to be\[
E_{0}^{\nu(1)}-E_{0}^{\nu(0)}=\frac{|\Omega_{\mathrm{c}}|^{2}}{\omega_{0}-(\sqrt{2}+1)\eta}.\]
 After equating the above to $2\eta$, the amplitude needed can be
obtained as\begin{equation}
\Omega_{\mathrm{c}}=\sqrt{2\eta\left[\omega_{0}-(\sqrt{2}+1)\eta\right]}.\label{eq:epsilon_c}\end{equation}

Comparing Eq.~\eqref{eq:epsilon_c} with Eq.~\eqref{eq:crit_cond},
we can observe that, in order to exhibit electromagnetically induced
transparency, the coupling amplitude $\Omega_{\mathrm{c}}$ would
have to be within a range such that it can simultaneously overcome
the spontaneous relaxation and yet prevent the system from being degenerate,
in addition to the requirement that $\omega_{q}$ be outside the critical
range indicated by Eq.~\eqref{eq:crit_Omega}. Without this range,
the dressed medium becomes unresponsive to the incident probe signal
and the absorption spectrum becomes flat.

If we draw the analogy of the lower dressed levels to the Zeeman sublevels
of the degenerate ground state~\cite{akulshin98}, then $\left|\mu_{0}\right\rangle $
and $\left|\nu_{0}\right\rangle $ can be deemed sharing the same
non-zero ``angular momentum.'' The remaining factor for deciding the
dressed medium to be transparent or absorbing depends on whether the
transition between the two sublevels is open or closed; in other words,
whether the transitions within the three-level system are non-cyclic
($\Lambda$-type) or cyclic ($\Delta$-type). Different from the usual
case, where this factor is determined by the hyperfine structure of
a particular atom, the dressed qubit system we discuss here has this
factor effectively determined by the dressed relaxation rates $\gamma_{\mu1}$
and $\gamma_{\mu\nu}$.

From Eqs.~\eqref{eq:gamma_mu1}-\eqref{eq:gamma_munu}, we observe
that unlike the usual multi-level SQUID systems, the relaxation rates
of the dressed qubit are tunable through the transformation angles
$\theta_{0}$ and $\theta_{1}$. In addition, the relaxation rate
$\gamma_{\mu\nu}$ between the lower levels is actually derived from
the dephasing rate $r_{\varphi}$ of the undressed qubit. The type
of transitions in the dressed qubit thus becomes exploitable since
the dephasing rates of the superconducting qubits are in general much
greater than their relaxation rates $r_{1}$ and the different dependences
on the transformation angles control whether $|\gamma_{\mu\nu}|>|\gamma_{\mu1}|$
or $|\gamma_{\mu\nu}|<|\gamma_{\mu1}|$.

When the qubit is closely resonant with the CPW resonator, or precisely,
$\omega_{q}$ is within the critical range between $\lambda_{\mathrm{C},-}$
and $\lambda_{\mathrm{C},+}$, the magnitude of $\gamma_{\mu\nu}$
is greater, under which the flipping processes of the populations
between the ground and the metastable states dominate over the excitation
process from the ground state to the excited state. The transition
between the ground and the metastable states is thus closed, effectively
degenerating the two levels and making the three-level system operate
in a $\Delta$-type setting. On the other hand, when the qubit is
off-resonant with the CPW resonator, the opposite condition $\gamma_{\mu1}>\gamma_{\mu\nu}$
is met and the lower two levels become sufficiently non-degenerate
that the transitions among the three levels cannot be considered cyclic.
The system then operates in the usual $\Lambda$-type setting that
electromagnetically induced transparency can take place.

\section{Numerical analysis\label{sec:implementations}}

We now study the two parts of the susceptibility by considering experimentally
accessible parameters. We first examine the dressed charge qubit,
which was experimentally realized in Refs.~\cite{wilson07,wilson09}.
We give the variation of the susceptibility against multiple parameters
and identity the effective ranges of the qubit level spacing for the
EIT and EIA. The cases for phase and flux qubits are discussed later
to show that the arguments for charge qubits can be applied to other
qubits, for tuning the susceptibility to different operating regimes.

\subsection{Charge qubit}

Without loss of generality, we now assume that the resonator frequency
is fixed at, e.g., $\omega_{0}/2\pi=7$~GHz. We consider the charge
qubit model with the following parameters: the qubit has a junction
energy $E_{J}/2\pi=2.6$~GHz and a charge energy in the GHz range
that we use to tune the qubit level spacing by varying the gate reduced
charge number $n_{g}$; the coupling coefficient between them is assumed
to be $\eta/2\pi=100$~MHz. The undressed relaxation and dephasing
times of the qubit are taken as $1/r_{1}=0.7\,\mu$s and $1/r_{2}=48$~ns,
respectively. The operating temperature is assumed to be 20~mK.

\begin{figure}
\includegraphics[bb=66bp 195bp 510bp 600bp,clip,width=9cm]{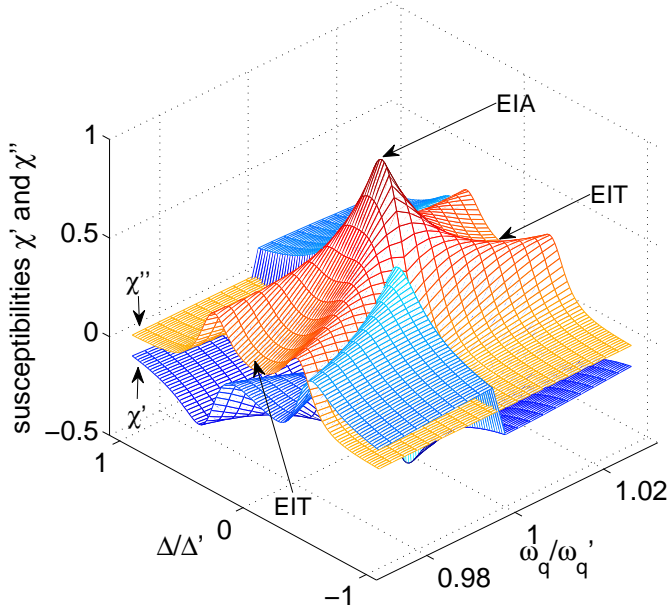}

\caption{(Color online) Real part $\chi'$ (blue) and imaginary part $\chi''$
(red or yellow) of the normalized susceptibility of the dressed charge
qubit versus the detuning $\Delta$ and the qubit level spacing $\omega_{q}$.
The detuning abscissa is normalized with respect to $\Delta'/2\pi=25$~MHz;
the qubit-level-spacing abscissa is normalized with respect to $\omega_{q}'/2\pi=3.5$~GHz.
This plot indicates the dispersion and absorption spectra of a probe
signal over different operating ranges.\label{fig:3D_suscept_charge}}

\end{figure}

Figure~\ref{fig:3D_suscept_charge} plots both the real $\chi'$
and the imaginary $\chi''$ parts of the susceptibility $\chi$ in
normalized units as a function of the normalized probe-signal detuning
$\Delta/\Delta'$ (with respect to the normalizing constant $\Delta'/2\pi=25$~MHz)
and the normalized qubit level spacing $\omega_{q}/\omega_{q}'$ (with
respect to the normalizing constant $\omega_{q}'/2\pi=3.5$~GHz)
near the resonance range of the dressed qubit. The strongly coupled
control field can achieve a coupling amplitude on the order of 10~MHz~\cite{wilson07};
the plot here assumes a value of $\Omega_{\mathrm{c}}/2\pi=12$~MHz.

For the imaginary part $\chi''$ (warm colored: red or yellow), we
are able to observe the absorption peaking at the resonant frequency
$\omega_{q}=\omega_{0}/2=2\pi\times3.5$~GHz and its immediate falloff
when $\omega_{q}$ is tuned slightly off-resonance. Along with the
attenuation of the magnitude is the symmetric splitting of the peak
about the zero detuning point $\Delta=0$ at either off-resonant side.
The zero-detuning point itself falls from its maximum value to its
local minimum along the path of off-resonance, indicating the switching
of the dressed medium from ``maximally absorbing'' to being transparent
to the probe signal.

\begin{figure}
\includegraphics[bb=52bp 145bp 525bp 614bp,clip,width=8.5cm]{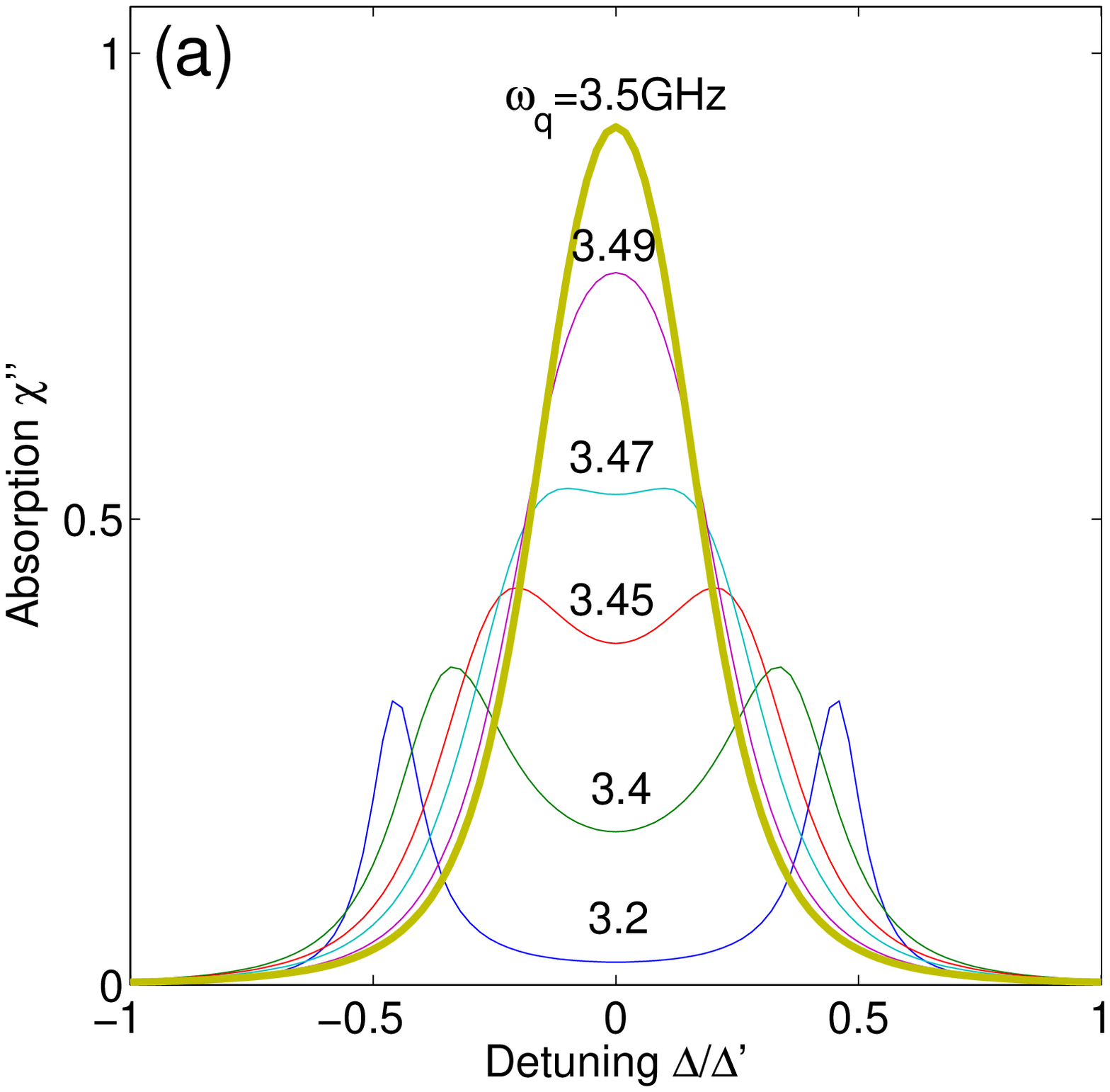}

\includegraphics[bb=40bp 148bp 525bp 611bp,clip,width=8.5cm]{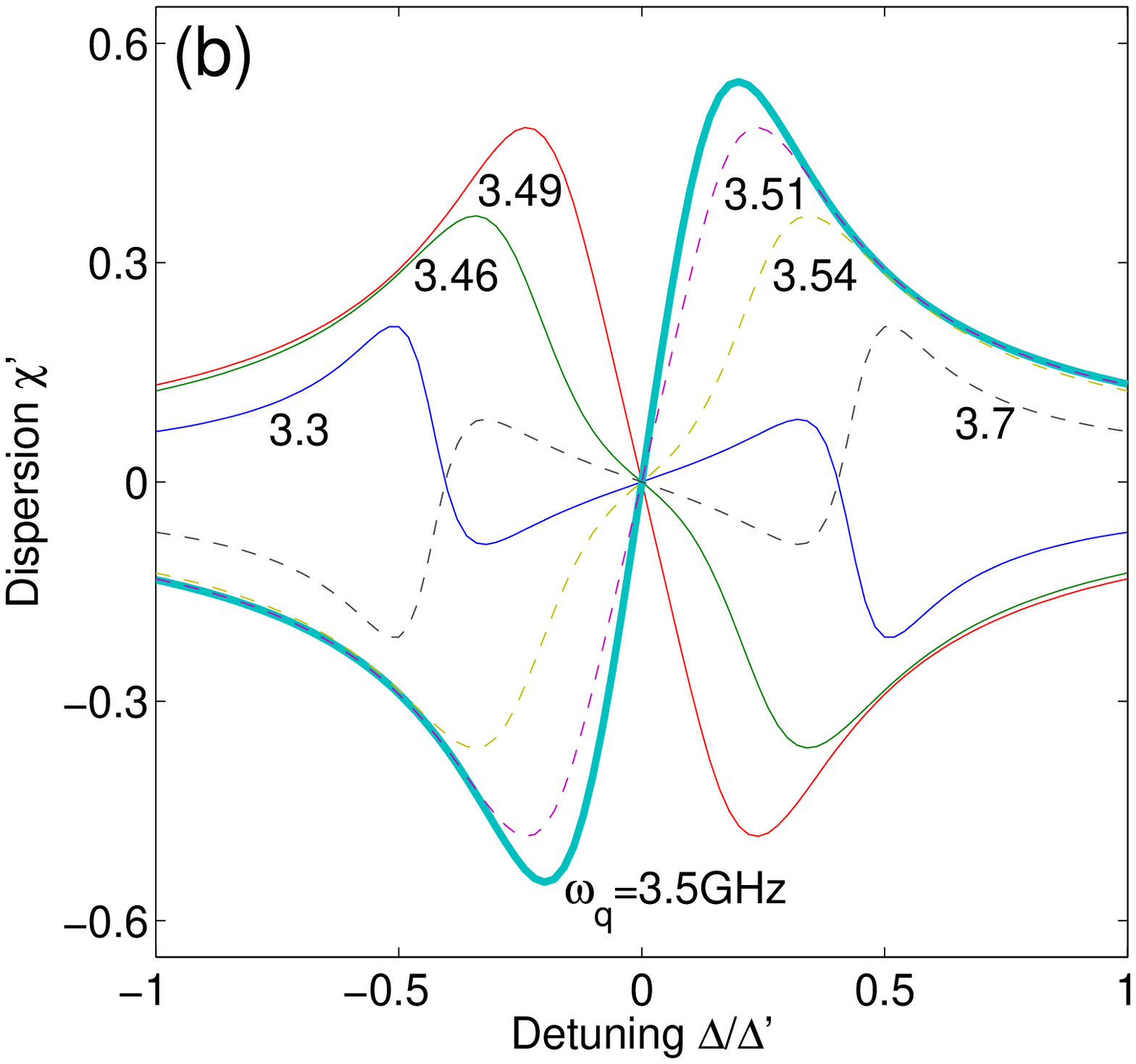}

\caption{(Color online) Normalized susceptibility spectra at typical values
of the qubit level spacing $\omega_{q}$ plotted versus the scaled
detuning $\Delta/\Delta'$: (a) the absorption $\chi^{\prime\prime}=\mathrm{Im}[\chi]$
and (b) the dispersion $\chi^{\prime}=\mathrm{Re}[\chi]$. The curves
that correspond to the resonant frequency $\omega_{0}/2=2\pi\times3.5$~GHz
are thickened and all the numbers indicate the value of $\omega_{q}$
taken (in unit of GHz) in both (a) and (b). Note that in (a) a dip
appears at the center of the curve about $\omega_{q}/2\pi\approx3.47$~GHz,
which indicates the \emph{switching of the dressed qubit from being
absorptive} ($>3.47$~GHz) \emph{to being transparent} ($<3.47$~GHz).
In (b), the curves that correspond to $\omega_{q}$ below $\omega_{0}/2$
are solid and those above $\omega_{0}/2$ are dashed. Note that the
solid and dashed curves with equal distance from the resonance $\omega_{0}/2$
are symmetric counterparts of each other. Their roles for positive
and negative dispersion across the range of the detuning $\Delta$
are exchanged above and below the resonance frequency.\label{fig:2D_suscept_charge}}

\end{figure}

The switching phenomenon is better illustrated in Fig.~\ref{fig:2D_suscept_charge}(a)
where the imaginary part $\chi''$ of the susceptibility is plotted
for various values of the spacing $\omega_{q}$. We note that the
susceptibility obtains a maximum value and a minimum half-width with
a Lorentzian shape when the dressed medium is resonant (the thickened
curve for $\omega_{q}/2\pi=3.5$~GHz). Following the detuning between
the qubit and the CPW resonator, the half-width starts to spread out
while the peak starts to dent. The switching occurs at the critical
value $\lambda_{\mathrm{C}}/2\pi=3.40$~GHz, according to the approximate
analytical solution of Eq.~\eqref{eq:crit_Omega}, where the qubit-to-resonator
coupling $\eta$ is not very large. A numerical estimate gives $\lambda_{\mathrm{C}}/2\pi\approx3.47$~GHz,
as can be seen from Fig.~\ref{fig:2D_suscept_charge}(a).

The plot of Fig.~\ref{fig:3D_suscept_charge} also shows that the
falloff of the absorption is even-symmetric about the resonance point
$\omega_{q}=\omega_{0}/2$, due to the fact that $\chi''$ is an even
function of the dressed relaxation rates $\gamma_{\mu1}$ and $\gamma_{\mu\nu}$,
which in turn are odd functions of the qubit spacing $\omega_{q}$.
This symmetry verifies the exchangeable roles of the near-degenerate
levels $\left|\mu_{0}\right\rangle $ and $\left|\nu_{0}\right\rangle $:
when $\omega_{q}<\omega_{0}/2$, $\left|\mu_{0}\right\rangle $ has
the lower eigenenergy of the dressed states and is regarded as the
ground state that couples to the probe signal, while $\left|\nu_{0}\right\rangle $
is regarded as the metastable state; when $\omega_{q}>\omega_{0}/2$,
$\left|\mu_{0}\right\rangle $ becomes the metastable level and $\left|\nu_{0}\right\rangle $
becomes the ground state.

In Fig.~\ref{fig:3D_suscept_charge}, for the real part $\chi'$
(blue colored), we can see the dispersion spectrum having an odd symmetry
around the zero probe detuning point $\Delta=0$, i.e. $\chi'(\omega_{q})|_{\Delta=0}=0$,
throughout the range of the qubit spacing $\omega_{q}$. Unlike $\chi''$,
this real part $\chi'$ of the susceptibility is an odd function of
the dressed relaxation rates, resulting in its odd symmetry of $\omega_{q}$
around the resonance point $\omega_{q}=\omega_{0}/2$, which is better
illustrated in Fig.~\ref{fig:2D_suscept_charge}(b). There, the dispersion
spectrum for various values of $\omega_{q}$ is plotted versus the
detuning $\Delta$. The one curve at resonance ($\omega_{q}/2\pi=3.5$~GHz)
is thickened and the ones that have their symmetric counterparts below
resonance are shown as dashed curves, from which we can observe the
switching from positive dispersion to negative dispersion of the dressed
medium discussed in Sec.~\ref{sec:tuning}.

\begin{figure}
\includegraphics[bb=30bp 212bp 545bp 625bp,clip,width=9cm]{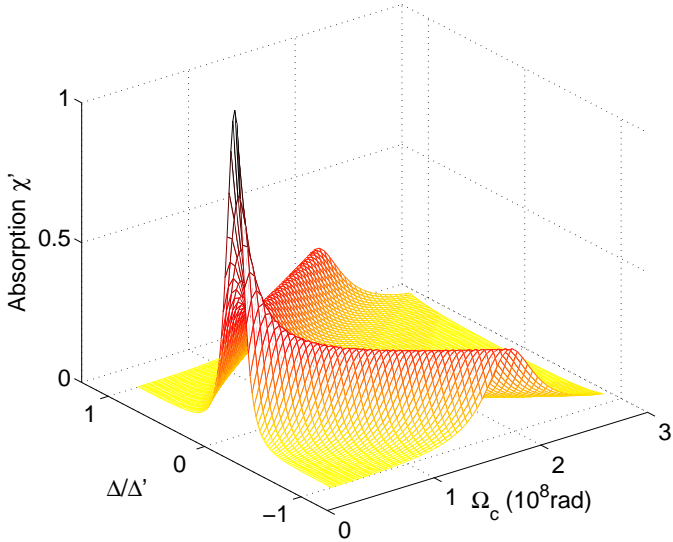}

\caption{(Color online) Plot of the normalized absorption spectrum $\chi^{\prime\prime}=\mathrm{Im}[\chi]$
versus the normalized detuning $\Delta/\Delta'$ and the coupling
amplitude $\Omega_{\mathrm{c}}$ to the control field.\label{fig:3D_suscept_epsilonc} }

\end{figure}

In addition to the spacing $\omega_{q}$, tuning the dressed qubit
from transparency to absorption also depends on the dressed system's
coupling strength $\Omega_{\mathrm{c}}$ to the control field since
this coupling strength directly affects the absorption (Cf. Eq.~\eqref{eq:scpt_im}).
Its influence is illustrated in Fig.~\ref{fig:3D_suscept_epsilonc}
where the absorption is plotted against $\Delta/\Delta'$ (over the
same range as in Fig.~\ref{fig:3D_suscept_charge}) and $\Omega_{\mathrm{c}}$
(from 4.5~MHz to 45~MHz) while the qubit spacing is held at a typical
value $\omega_{q}/2\pi=3.4$~GHz. We can notice the single peaking
at the lower end of the coupling to the control field, where the dressed
medium is weakly driven by the control field and exhibits a population
trapping of the probe signal. Towards higher values of the coupling,
the excited state of the dressed three-level system is sufficiently
detuned from the probe signal that it starts to exhibit the transparency
effect. Then similar to ordinary $\Lambda$-type atoms, the dressed
qubit has its twin absorption peaks further apart when the coupling
strength is increased.

\subsection{Phase qubits}

We now examine our general theory of the tunable transparency and
absorption effects on other superconducting quantum circuit systems.

\begin{figure}
\includegraphics[bb=56bp 170bp 523bp 579bp,clip,width=8.5cm]{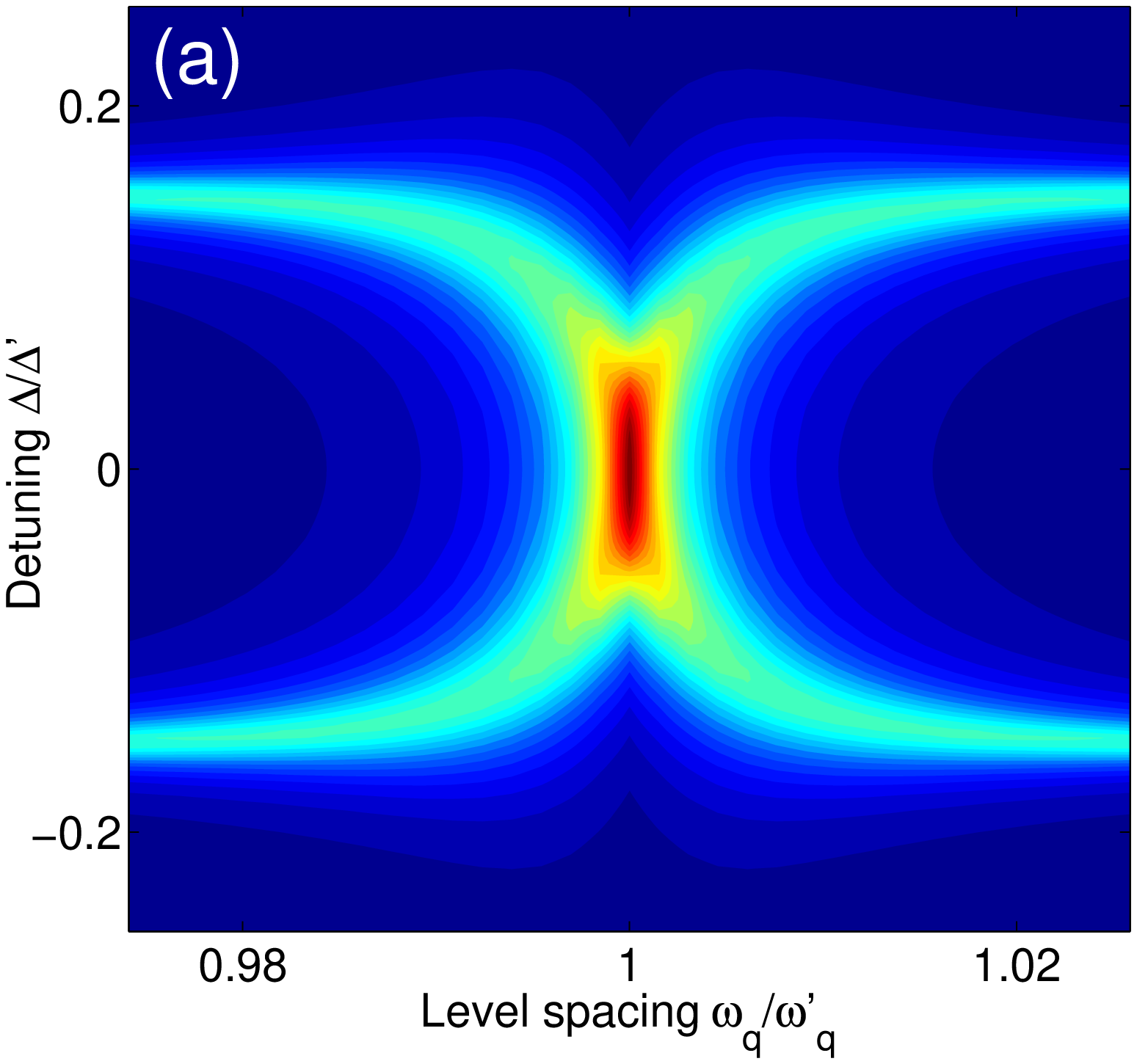}

\includegraphics[bb=50bp 170bp 528bp 583bp,clip,width=8.5cm]{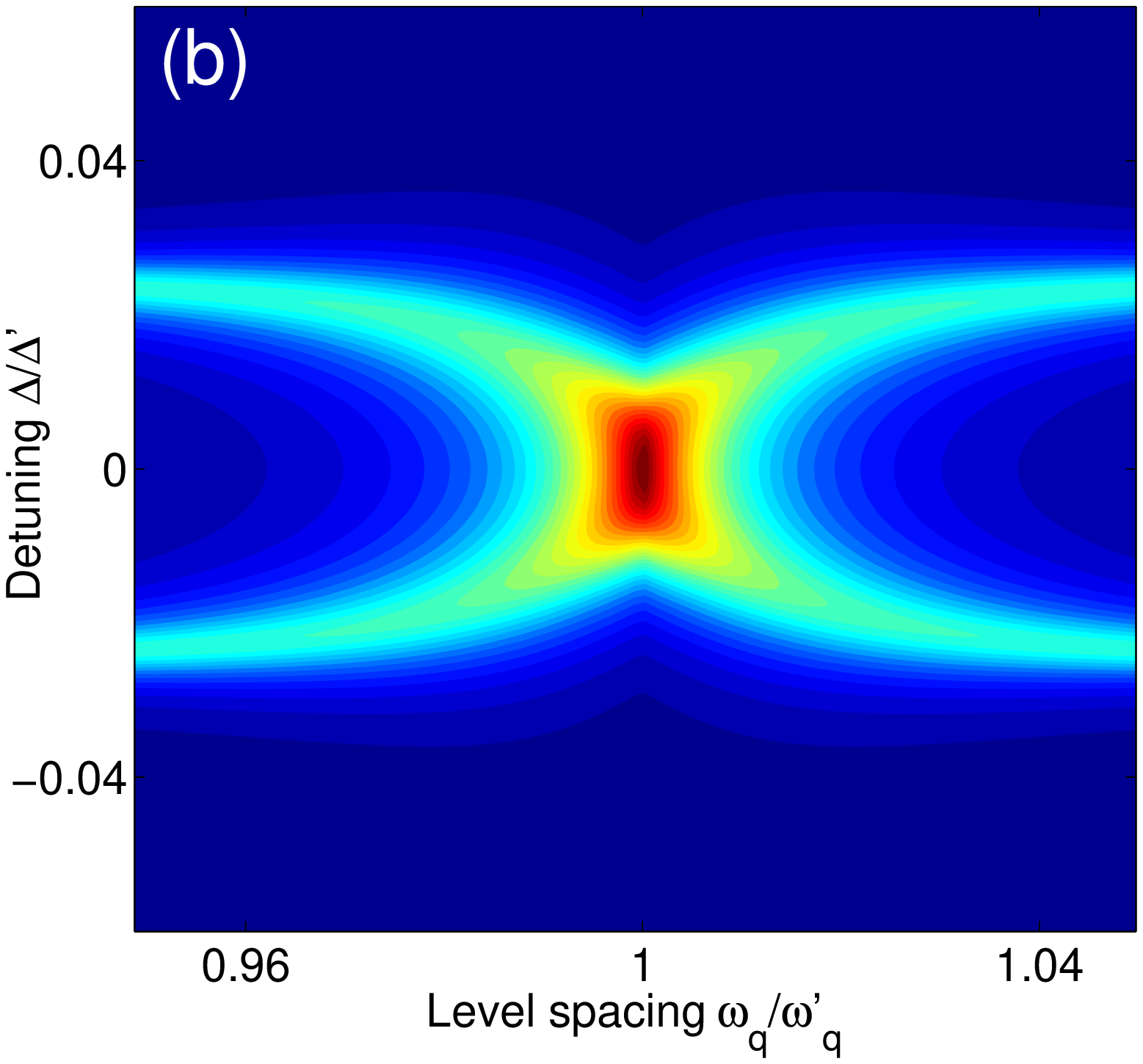}

\caption{(Color online) Contour plots of the normalized absorption spectra
$\chi''$ versus the normalized detuning $\Delta/\Delta'$ on the
vertical axis and the normalized qubit level spacing $\omega_{q}/\omega_{q}'$
on the horizontal axis: (a) for a phase qubit and (b) for a flux qubit.
The red end of the color spectrum (i.e., the centers of both contour
plots) indicates peak values of the absorption, where the qubit is
maximally dressed with a rotation angle $\theta_{0}=\pi/2$. The deep
blue end of the color spectrum for $\Delta/\Delta'=0$ (i.e., the
middle sections of the left and right borders of both plots) indicates
the minimum values of the absorption, where the qubit is dressed with
a rotation angle $\theta_{0}<\pi/2$. Namely, maximum transmission
due to EIT occurs at those blue ends.~\label{fig:contour_suscept}}

\end{figure}

For a phase qubit, we adopt the experimental parameters of Ref.~\cite{cleland09}:
the CPW resonator has a frequency $\omega_{0}/2\pi=6.57$~GHz; the
coupling strength between the resonator and the qubit is fixed at
$\eta/2\pi=19$~MHz; the undressed relaxation and dephasing times
of the qubit are, respectively, 650~ns and 150~ns. The coupling
strength to the control signal is assumed to be $\Omega_{\mathrm{c}}/2\pi=3.85$~MHz.
The operation temperature is held at 25~mK. The normalized absorption
spectrum is plotted as a contour plot versus the normalized qubit
level spacing $\omega_{q}/\omega_{q}'$ with respect to the charge
qubit case ($\omega_{q}$ from 3.2~GHz to 3.37~GHz; normalizing
constant $\omega_{q}$ the same as that of the charge qubit) and the
normalized probe detuning $\Delta/\Delta'$ ($\Delta$ from -6.5~MHz
to 6.5~MHz; normalizing constant $\Delta'$ the same as that of the
charge qubit) in Fig.~\ref{fig:contour_suscept}(a).

Similar to the case of a charge qubit in the last subsection, the
peak at the center of the contour falls off symmetrically as the qubit
spacing is tuned off the resonant frequency $\omega_{q}/2\pi=3.285$~GHz.
The peak also splits symmetrically around the zero detuning $\Delta=0$,
and enters from the absorption region into the transparency region
at the critical frequency $\lambda_{\mathrm{C}}/2\pi=3.266$~GHz
according to Eq.~\eqref{eq:crit_Omega}. The half-widths of the absorption
peaks decrease along with the fall off of the magnitude.

\subsection{Flux qubits}

For a flux qubit, we adopt the experimental parameters of Ref.~\cite{tsai08}:
the CPW resonator has oscillating frequency $\omega_{0}/2\pi=9.907$~GHz;
the qubit-resonator coupling strength is $\eta/2\pi\approx100$~MHz;
the undressed relaxation and dephasing times of the qubit are, respectively,
1.9~$\mu$s and 1~$\mu$s \cite{yoshihara06}. The coupling strength
to the control signal is set to $\Omega_{\mathrm{c}}/2\pi=0.63$~MHz.
The operation temperature is 50~mK. The normalized absorption spectrum
of the dressed flux qubit is also plotted as a contour plot versus
the normalized qubit level spacing $\omega_{q}/\omega_{q}'$ with
respect to the charge qubit case ($\omega_{q}$ from 4.7~GHz to 5.2~GHz;
normalizing constant $\omega_{q}'$ the same as that of the charge
qubit) and the normalized probe detuning $\Delta/\Delta'$ ($\Delta$
from -1.5~MHz to 1.5~MHz; normalizing constant $\Delta'$ the same
as that of the charge qubit) in Fig.~\ref{fig:contour_suscept}(b).

Similar operating regions of transparency and absorption can be observed
in this flux qubit system compared to the other qubit systems discussed
before, except for the scale of variations. For instance, comparing
Figs.~\ref{fig:contour_suscept}(a) and (b), we see that the fall-off
of the magnitude along the zero detuning is relatively slower in the
flux qubit case due to the slower dephasing time of the undressed
flux qubit. The switching occurs at the critical frequency $\lambda_{\mathrm{C}}/2\pi=4.854$~GHz.
In this case, the absorption peaks also split out slower and have
narrower half-widths.

\section{Conclusion\label{sec:conclusion}}

We have proposed a method to realize the effects of both EIT and EIA
on superconducting quantum circuits using dressed states derived from
the coupling between an arbitrary type of two-level superconducting
junction qubits and a coplanar waveguide resonator. The use of dressed
states alleviates the need to maintain a multi-level structure of
the Josephson devices and gives rise to tunable relaxation rates between
the energy levels. The tunable structure of the levels leads to the
switching between EIT and EIA, which depends on the variable qubit
level spacing and is associated with the open or closed transition
structure and the hyperfine degeneracy of the dressed three-level
system.

Our investigation demonstrates another example of nonlinear optical
phenomena implementable on superconducting quantum circuits. We can
also see that the special characteristics of Josephson junction devices,
the externally controllable Josephson coupling energy in this case,
could bring new perspectives to the study of quantum optics where,
for example, the many parameters are usually fixed for the particular
type of atom studied and the cavity QED system that surrounds it.
The switching between EIT and EIA might have important applications
for the control of superconducting circuits and for quantum information
transfer in these systems.
\begin{acknowledgments}
We thank Prof. C. P. Sun for discussions. FN was supported in part
by the National Security Agency, Laboratory of Physical Sciences,
Army Research Office, National Science Foundation Grant No. 0726909,
and JSPS-RFBR Contract No. 09-02-92114, MEXT Kakenhi on Quantum Cybernetics,
and FIRST (Funding Program for Innovative R\&D on S\&T). YXL acknowledges
support from the National Natural Science Foundation of China under
No. 10975080.
\end{acknowledgments}
\appendix

\section{Condition of local extrema for the susceptibility}

The two roots of Eq.~\eqref{eq:Delta_root} that correspond to local
extrema of the dispersion spectrum must satisfy\[
(\gamma_{\mu1}+\gamma_{\mu\nu})\zeta_{c}\sqrt{\zeta_{c}^{2}+\gamma_{\mu1}\gamma_{\mu\nu}}-\gamma_{\mu\nu}(\zeta_{c}^{2}+\gamma_{\mu1}\gamma_{\mu\nu})>0,\]
which is equivalent to \[
(\gamma_{\mu1}+\gamma_{\mu\nu})\zeta_{c}>\gamma_{\mu\nu}\sqrt{\zeta_{c}^{2}+\gamma_{\mu1}\gamma_{\mu\nu}}\;.\]
When squaring the two sides, the above inequality implies\[
\gamma_{\mu1}^{2}\zeta_{c}^{2}+2\gamma_{\mu1}\gamma_{\mu\nu}\zeta_{c}^{2}-\gamma_{\mu1}\gamma_{\mu\nu}^{3}>0\,.\]
For $\zeta_{c}^{2}>-\gamma_{\mu1}\gamma_{\mu\nu}$, with the relaxation
rates taking values $\gamma_{\mu1}<0$ and $\gamma_{\mu\nu}>0$ from
Eqs.~\eqref{eq:gamma_mu1}-\eqref{eq:gamma_munu}, we have\begin{eqnarray*}
 &  & \gamma_{\mu1}\left\{ (\gamma_{\mu1}+2\gamma_{\mu\nu})\zeta_{c}^{2}-\gamma_{\mu\nu}^{3}\right\} \\
 & > & \gamma_{\mu1}\left\{ -(\gamma_{\mu1}+2\gamma_{\mu\nu})\gamma_{\mu1}\gamma_{\mu\nu}-\gamma_{\mu\nu}^{3}\right\} \\
 & = & -\gamma_{\mu1}\gamma_{\mu\nu}(\gamma_{\mu1}+\gamma_{\mu\nu})^{2}>0,\end{eqnarray*}
 so the real roots exist under this condition.

\end{document}